\documentclass[aps, prd,twocolumn,groupedaddress]{revtex4-2}

\usepackage{xcolor}
\usepackage{orcidlink}

\usepackage{dcolumn}
\usepackage{bm}
\usepackage{hyperref}
\usepackage{amsmath}
\usepackage{graphicx}    
\usepackage{subcaption}  
\usepackage{cancel}
\usepackage[left]{lineno}

\newcommand{\gpcr}{\texttt{sub-sample systematics}}

\newcommand{\gpcrB}{\texttt{Sub-sample systematics}}

\newcommand{\regularsys}{\texttt{1st order systematics}}

\newcommand{\ngal}{$n_{\mathrm{gal}}/\bar{n}_{\mathrm{gal}}$}

\newcommand{\nobs}{$\rho_{\mathrm{obs}}$}
\newcommand{\wsys}{$w_\mathrm{sys}$}
\usepackage{xcolor}
\newcommand{\hui}[1]{\textcolor{black}{#1}}

\newcommand{\huip}[1]{\textcolor{black}{#1}}

\usepackage{tikz}
\usetikzlibrary{shapes.geometric, positioning} 

\begin{document}

\preprint{APS/123-QED}

\title{
  \texorpdfstring{Imaging systematics induced by galaxy sub-sample fluctuation: \\ new systematics at second order}{The Inhomogeneous Distribution of Galaxy Bias and Redshift: The sub-sample imaging systematics at second order}
}

\author{Hui Kong\,\orcidlink{0000-0001-8731-1212}}
\email{hkong@ifae.es}
 \affiliation{Institut de F\'{i}sica d’Altes Energies (IFAE), The Barcelona Institute of Science and Technology, Campus UAB, 08193 Bellaterra Barcelona, Spain} 
 
\author{Nora Elisa Chisari\,\orcidlink{0000-0003-4221-6718}}
\affiliation{Institute for Theoretical Physics, Utrecht University, Princetonplein 5, 3584 CC, Utrecht, The Netherlands.}

\author{Boris Leistedt\,\orcidlink{0000-0002-3962-9274}}
\affiliation{Department of Physics, Imperial College London, Blackett Laboratory, Prince Consort
Road, London SW7 2AZ, UK}

\author{Eric Gawiser\,\orcidlink{0000-0003-1530-8713}}
\affiliation{Department of Physics and Astronomy, Rutgers, the State University of New Jersey, Piscataway, NJ 08854, USA}

\author{Martin Rodríguez-Monroy\,\orcidlink{0000-0001-6163-1058}}
\affiliation{Instituto de Fisica Teorica UAM/CSIC, Universidad Autonoma de Madrid, 28049 Madrid, Spain}

\author{Noah Weaverdyck\,\orcidlink{0000-0001-9382-5199}}
\affiliation{Lawrence Berkeley National Laboratory, 1 Cyclotron Road, Berkeley, CA
94720, USA}

\author{The LSST Dark Energy Science Collaboration}

\date{\today}

\begin{abstract}
Imaging systematics refers to the inhomogeneous distribution of a galaxy sample caused by varying observing conditions and astrophysical foregrounds. Current mitigation methods correct the galaxy density fluctuations \ngal\, caused by imaging systematics assuming that all galaxies in a sample have the same \ngal. Under this assumption, the corrected sample cannot perfectly recover the true correlation function. We name this effect \gpcr. For a galaxy sample, even if its overall sample statistics (redshift distribution $n(z)$, galaxy bias $b(z)$), are accurately measured, $n(z), b(z)$ can still vary across the observed footprint. It makes the correlation function amplitude of galaxy clustering higher, while correlation functions for galaxy-galaxy lensing and cosmic shear do not have noticeable change. Such a combination could potentially degenerate with physical signals on small angular scales, such as the amplitude of galaxy clustering, the impact of neutrino mass on the matter power spectrum, etc. \gpcr\, cannot be corrected using imaging systematics mitigation approaches that rely on the cross-correlation signal between imaging systematics maps and the observed galaxy density field. In this paper, we derive formulated expressions of \gpcr, demonstrating its fundamental difference with other imaging systematics. We also provide several toy models to visualize this effect. Finally, we discuss a potential method to estimate and mitigate \gpcr\, by forward modeling its behavior using Synthetic Source Injection. 
\end{abstract}

\maketitle

\section{Introduction}

The Dark Energy Science Collaboration (DESC) \footnote{\href{https://lsstdesc.org/}{DESC home page}} will obtain state-of-the-art, robust cosmology constraints by taking advantage of the data from the Vera Rubin Observatory’s Legacy Survey
of Space and Time (LSST) \footnote{\href{https://rubinobservatory.org/}{Rubin Observatory home page}}. The high-level science requirement of DESC \cite{mandelbaum2018lsst} is in the form of Dark Energy Task Force (DETF) Figure of Merit (FoM) \cite{albrecht2006report}, suggesting large improvements \cite{font2014desi} over stage III \cite{albrecht2006report} galaxy surveys. This ambitious goal must be supported by a comprehensive understanding of potential systematics that could introduce bias into the measurement. This brings increased scrutiny to aspects rarely considered in Stage III surveys. In the theoretical aspect, the effects of baryonic feedback \cite{truttero2024baryon,troster2022joint, zennaro20241}, non-linear galaxy bias \cite{nicola2024galaxy} come into focus. On the observational side, the effects of blending \cite{levine2024galaxy,mendoza2024blending,biswas2024madness} have become increasingly important. There are numerous other ideas that cannot be fully enumerated here \cite{shah2025oracle,merz2024deepdisc,zeghal2024simulation,armstrong2024little}. In this work, we raise a new idea concerning \textit{imaging systematics}, which requires a comprehensive investigation to ensure the robustness of future cosmological measurements from DESC.

When we select a galaxy sample with a selection function defined by the astronomical source's color, shape, etc, we want the sample to be homogeneously distributed. However, the observed properties for a given galaxy are not strictly the same as its true properties. When applying a selection function defined by the galaxies` observed properties, this phenomenon creates inhomogeneity: the selected galaxy sample is not homogeneously distributed across the full footprint. This effect is defined as \textit{imaging systematics}. A comprehensive understanding of how systematics affect cosmological observables is needed to meet the requirements of the DESC survey's science goal \cite{lochner2018optimizing, hang2024impact, moskowitz2023improved, prat2022catalog,zhang2023impact}. A well studied \cite{ho2012clustering,ross2011ameliorating,elvin2018dark,wagoner2021linear,rodriguez2022dark, chaussidon2022angular,2020MNRAS.495.1613R,johnston2021organised,yan2024kids,kong2020removing} effect caused by  \textit{imaging systematics} is the fluctuations of \textit{relative galaxy number densities} (referred to as \ngal\, below). The observed galaxy density (referred to as \nobs\,  below) is correlated with certain imaging systematics maps because such maps change the number density of the galaxies that enter the selection function. 

The \ngal\, variations due to imaging systematics are degenerate with changes in galaxy density due to variations in cosmological parameters. Various methods have been developed to separate the two types of variations and obtain accurate cosmological inference. Some of these methods assume that \ngal\, is linearly related to the imaging systematics maps \cite{ho2012clustering,ross2011ameliorating,elvin2018dark,wagoner2021linear,rodriguez2022dark}, others take a non-linear relationship \cite{chaussidon2022angular,2020MNRAS.495.1613R,rodriguez2022dark,johnston2021organised,yan2024kids}. Despite differences in the detailed implementation, all methods share the same philosophy:

\textit{The imaging systematics mitigation is based on the cross-correlation signal of the observed galaxy density field \nobs\, and the imaging systematics maps.}

Some of these methods apply corrections directly to the measured correlation functions (or power spectrum) \cite{ho2012clustering,ross2011ameliorating}, while others produce imaging systematics weights (which we will call \wsys) to each galaxy \cite{elvin2018dark,wagoner2021linear,rodriguez2022dark,chaussidon2022angular,2020MNRAS.495.1613R,johnston2021organised,yan2024kids,kong2020removing} based on its sky position. \wsys\, can also be applied to the randoms used for measuring correlation function instead of the observed galaxies (e.g. \cite{yan2024kids} produces `organized randoms' based on the self-organizing maps). \wsys\, ensures that the fluctuations of the \ngal\, trend with weighted \nobs\, against all imaging systematics maps, are consistent with mocks that do not have any imaging systematics. For convenience, we only consider the method that produces  imaging systematics weights. Other methods can be mathematically transformed to this method \cite{weaverdyck2021mitigating}. With the properly measured imaging systematics weight \wsys, the observed galaxy density field is:
\begin{equation}
    \rho_\mathrm{corrected} = w_\mathrm{sys} \rho_\mathrm{obs}
    \label{equ:general-sys}
\end{equation}
 \wsys\, changes with sky positions, usually in the form of \textsc{healpix} \cite{gorski2005healpix} pixels. Such simplification gives rise to a question: Does an optimal retrieval of $w_\mathrm{sys}$ for \nobs\, guarantee that the corrected galaxy density field $\rho_\mathrm{corrected}$ is identical to the true galaxy density field $\rho_\mathrm{truth}$?

One obvious simplification of equation \ref{equ:general-sys} is that it assumes all galaxies within the sample respond to all imaging systematics in the same way, and $w_\mathrm{sys}$ is the same for all galaxy samples. Such an assumption is not strictly true: though galaxies in the same sample share similar properties, they still have variations in their intrinsic properties like color, shape, luminosity, etc. We refer to these variations as `types'. Such difference means that different types of galaxies respond to imaging systematics differently. For example, the bright and faint galaxies have different sensitivity to the noise level of the image (depth), thus they would have different \ngal\, trends against the depth map \cite{kong2024forward}.

We present a detailed study of this phenomenon, and introduce the concept of \gpcr: the change in the composition of different galaxy types causes a spatial variation of the redshift distribution $b(z, \textbf{sys})$ and galaxy bias $b(z, \textbf{sys})$. \textbf{sys} is the systematics map value at different sky positions \textbf{sys}($\Omega$). We explore the systematics effects assuming accurate overall sample statistics, including accurate global \wsys, the overall redshift distribution $n(z)$, and the overall galaxy bias $b(z)$. We only assume that the true distribution of different types of galaxies is inhomogeneous.

This work is organized as follows: 

Section \ref{sec:theory} starts from the traditional imaging systematics expressions and expands them to the theoretical formalisms of \gpcr. We further discuss some properties of this effect. Section \ref{sec:example-case-ebv} uses a toy model assuming a wrong dust extinction map, and presents a numerical way to estimate this effect. Section \ref{sec:three-toy-models} introduces some toy models to address this effect analytically, discussing its impact on galaxy clustering, galaxy-galaxy lensing, and cosmic shear. Section \ref{sec:clumping-cosmology} discusses its impact on cosmological measurements. Section \ref{sec:ssi} discusses estimating the effect with synthetic source injection. Section \ref{sec:conclusion} provides a summary and conclusion of this work. 

\section{Theory}
\label{sec:theory}

In this section, we develop the theoretical framework underlying \gpcr. We begin by reviewing commonly used formalisms for imaging systematics and extend them to galaxy samples composed of multiple sub-samples affected by varying imaging conditions (Section \ref{sec:all-gals}). We then demonstrate that \gpcr\, arise naturally from one of the terms in this extended formalism, effectively altering the composition of galaxy pair counts across different angular separations (Section \ref{sec:gpcr}). 

\subsection{Imaging Systematics Background Knowledge}
\label{sec:all-gals}

In imaging systematics mitigation modeling, the observed galaxy density field is considered as \cite{elvin2018dark, ross2011ameliorating, kalus2019map, alonso2019unified, nicola2020tomographic}:

\begin{equation}
    \rho_\mathrm{obs} = \rho_\mathrm{no\,sys} \prod_{i} (1+\mathcal{F}({\textbf{sys}^{i}}))
    \label{equ:linear-contamin1}
\end{equation}

Here $\textbf{sys}^i$ are the systematics values in survey property maps, e.g. dust extinction \cite{schlegel1998maps}, size of the point spread function (PSF) \cite{jarvis2021dark}, etc. We use superscript $i$ to notate different systematics maps. $\mathcal{F}$ is a function that reflects how $\textbf{sys}^i$ changes the observed galaxy density. $\rho_{obs}$ is the observed galaxy density field. $\rho_\mathrm{no\,sys}$ represents the galaxy density field observed without any imaging systematics. Note that in this model, we ignore \textit{additive systematics}, which is mainly contributed by stellar contamination and can be modeled separately. Imaging systematics can change the mean observed density $\bar{\rho}_\mathrm{obs}$. For example, galaxies could be lost due to the noise in the images, reducing the number density. When the mean of $\mathcal{F}$ is not 0, the mean galaxy density is shifted. In a realistic situation, we do not consider $\rho_\mathrm{no\,sys}$ as the `truth catalog'. Instead, we just need a homogeneously distributed catalog with good knowledge of the redshift distribution $n(z)$ and galaxy bias $b(z)$. Under these considerations, we define the `truth catalog' to have the same mean number density as the `observed catalog'. $\mathcal{F}$ is re-defined to $f$, where $1+f$ is proportional to $1+\mathcal{F}$, and $f$ has a mean of 0. 

\begin{equation}
    \rho_\mathrm{obs} = \rho_\mathrm{truth} \prod_{i} (1+f({\textbf{sys}^{i}}))
    \label{equ:linear-contamin2}
\end{equation}

$\rho_\mathrm{truth}$ is not the same as the galaxy density field without any imaging systematics $\rho_\mathrm{no\,sys}$. Rather, it can be viewed as a re-selection of galaxies in $\rho_\mathrm{obs}$. This hypothetically selected sample is homogeneously distributed against all imaging systematics and has the same number density as $\rho_\mathrm{obs}$. In this particular case, we can set the mean of $f(\textbf{sys}^i)$ to 0. We denote the mean as a bar operator: 

\begin{equation}
     \bar{f}(\textbf{sys}^i) = 0,\, \bar{\rho}_\mathrm{obs} = \bar{\rho}_\mathrm{truth}
\end{equation}

\subsection{Imaging systematics with multiple sub-samples}
In a more realistic configuration, a galaxy sample comprises galaxies with different redshift, luminosity, etc. The defined galaxy sample can be seen as an ensemble of galaxies with similar properties. However, one type of galaxy in this ensemble could be different from another due to having a different intrinsic flux, color and shape. We consider the selected galaxy sample to be an ensemble of galaxies $k$ with various properties. Each $k$ responds to survey property maps differently. Equation \ref{equ:linear-contamin2} should be modified to: 

\begin{equation}
    \rho_{\mathrm{obs}} = \sum_k {\rho_{\mathrm{truth},k}} \prod_i (1 + f_{k} (\textbf{sys}^i))
    \label{equ:rho-obs-multi}
\end{equation}

Each $f_k(\textbf{sys}^i)$ still has a mean of 0. 
$\bar{\rho}_{\mathrm{truth}, k}$ is defined as the mean density of the galaxy sub-sample $k$. 

\begin{equation}
    \bar{\rho}_{\mathrm{truth}, k} = \frac{N_{\mathrm{obs},k}}{N_{\mathrm{obs}, \mathrm{total}}}\cdot \bar{\rho}_{\mathrm{obs}}
\end{equation}

Similarly to the discussion in Section \ref{sec:all-gals}, $\rho_{\mathrm{truth}, k}$ represents a galaxy sample that is homogeneously distributed against any imaging systematics, and has the same redshift and bias distribution as the observed galaxy sample. It is not the composition of the galaxy sub-sample $k$ without any systematics. Under this definition, we have $\bar{f}_k(\textbf{sys}^i)=0$.

We define the fraction of $\rho_{\mathrm{truth},k}$ in the whole sample is $h_k$:

\begin{equation}
    h_k = \bar{\rho}_{\mathrm{truth}, k}/\bar{\rho}_\mathrm{obs}
    \label{equ:fraction_hk}
\end{equation}

\subsubsection{Galaxy correlation function}
The relative galaxy density fluctuation is:

\begin{equation}
    \delta_\mathrm{obs} = \frac{\rho_\mathrm{obs}}{\bar{\rho}_\mathrm{obs}} - 1
\end{equation}

We expand it based on Equation \ref{equ:rho-obs-multi} and \ref{equ:fraction_hk} \hui{in terms of $f_k$}, 
and preserve it to the second order:

\begin{multline}
    \delta_{\mathrm{obs}} = \sum_{k} h_{k} [ \delta_{\mathrm{truth}, k} + \sum_i f_k(\textbf{sys}^i) + \\
    \delta_{\mathrm{truth}, k} \sum_i f_k(\textbf{sys}^i ) + \sum_{i>j} f_k(\textbf{sys}^i ) \cdot f_k(\textbf{sys}^j) + O(3)]
\end{multline}

$\delta_{\mathrm{truth}, k}$ is uncorrelated with any imaging systematics $f_{k\prime}$. Any density distribution on a sphere can be expressed with spherical harmonics:
\begin{equation}
\delta = \sum_{\ell m} a_{{\ell m}}Y_{{\ell m}}(\theta, \phi)
\end{equation}

Given a redshift shell, $a_{{\ell m}}$s of a galaxy distribution are only constrained by:
\begin{equation}
C_{\ell}^\mathrm{truth} = \frac{\sum_{m} a^{\mathrm{truth}}_{\mathrm{\ell m}}a^{*,\mathrm{truth}}_{\mathrm{\ell' m'}}\delta_{\ell \ell'} \delta_{mm'}}{2\ell+1} 
\end{equation}

The $a_{\ell m}^{\mathrm{truth}}$s can have arbitrary phases in ($\theta$, $\phi$) space. Due to this randomness, $a_{\ell m}^{\mathrm{truth}}$s from the true galaxy distribution is unlikely to be correlated with the $a^{i}_{\ell m}$ phases produced by a survey property map $i$. Terms like $\langle \delta_{\mathrm{truth}, k},f_{k}(\textbf{sys}^i)\rangle$ are expected to be 0. There are cases where Large-Scale-Structure signal leaks into the systematics maps like the Cosmic Infrared Background leaking into dust extinction map \cite{chiang2019extragalactic}. Such a situation is rare and can be properly handled with a corrected map \cite{chiang2023corrected}. We ignore this case in our modeling. 

For convenience, we denote: 

\begin{equation}
    f_k(\textbf{sys}^i) = f_k^i
    \label{equ:f_k-i}
\end{equation}

The correlation function of $\delta_{\mathrm{obs}}$ is:
\begin{multline}
    \langle \delta_{\mathrm{obs}},\delta_\mathrm{obs} \rangle = \sum_{kk\prime} h_{k}h_{k\prime}\langle \delta_{\mathrm{truth}, k},\delta_{\mathrm{truth}, k\prime}\rangle \\+\sum_{kk\prime}h_{k} h_{k\prime} \langle \sum_i f_k^i, \sum_{i\prime} f_{k\prime}^{i'} \rangle +\\
    2\sum_{kk\prime}h_kh_{k'}\langle\sum_{i>j} f_k^i\cdot f_k^j, \sum_{i\prime} f_{k\prime}^{i'}\rangle
    +\\
    \sum_{kk\prime}h_kh_{k'}\langle\sum_{i>j} f_k^i\cdot f_k^j, \sum_{i\prime > j\prime} f_{k\prime}^{i'} \cdot f_k^{j'}\rangle +\\
    2\sum_{kk\prime}h_kh_{k'}\langle\sum_{i>j>m} f_k^i\cdot f_k^j \cdot f_k^m, \sum_{i'} f_{k\prime}^{i'}\rangle
    + \\
    \sum_{kk\prime} h_{k}h_{k\prime}\langle \delta_{\mathrm{truth}, k} f_{k\prime}^i, \delta_{\mathrm{truth}, k\prime} f_{k'}^{i'}\rangle +\\
    \cancel{2\sum_{kk\prime} h_{k} h_{k\prime} \langle \sum_i f_k^i \delta_{\mathrm{truth}, k}, \delta_{\mathrm{truth},k\prime}\rangle} +\\ 
    + \cancel{2\sum_{kk\prime} h_{k} h_{k\prime} \langle \sum_i f_k^i \delta_{\mathrm{truth}, k}, f_{k\prime}^{i'}\rangle}
    \label{equ:long-version}
\end{multline}

To better visualize the equation above, we simplify the model as having only one imaging systematics $f_k$. In Appendix \ref{apdx:B} and \ref{apdx:C}, we take multiple systemtics maps $i$ into consideration. 
Here, the observed galaxy density field is:
\begin{equation}
    \delta_{\mathrm{obs}} = \delta_{\mathrm{truth}} + \sum_k f_k + \sum_k \delta_{\mathrm{truth}, k} f_k 
\end{equation}

We use the one systematics map configuration in future modeling. However, all our conclusions apply to the multiple imaging systematics maps scenario. The simplified correlation function is: 

\begin{multline}
    \langle \delta_{\mathrm{obs}},\delta_{\mathrm{obs}} \rangle = \langle \delta_{\mathrm{truth}},\delta_{\mathrm{truth}}\rangle \\
    +\sum_{kk\prime}h_{k} h_{k\prime} \langle f_k, f_{k\prime}\rangle \\ + 
    \sum_{kk\prime} h_{k}h_{k\prime}\langle \delta_{\mathrm{truth}, k} f_k, \delta_{\mathrm{truth}, k\prime} f_{k\prime}\rangle +\\
    \cancel{2\sum_{k} h_{k} \langle f_k \delta_{\mathrm{truth}, k}, \delta_{\mathrm{truth}}\rangle} +\\ 
    + \cancel{2\sum_{kk\prime} h_{k} h_{k\prime} \langle f_k \delta_{\mathrm{truth}, k}, f_{k\prime}\rangle}
    \label{equ:one-term-xi}
\end{multline}

We cross out terms that are 0. Appendix \ref{apdx:A} provides a proof for these terms being 0. Imaging systematics mitigation methods use \ngal\, to estimate the contribution of imaging systematics. \ngal\, is used both for mitigation and validation of imaging systematics. A detectable trend in \ngal\, is equivalent to having a signal in the cross-correlation of the systematics map and the observed galaxy density field. The observed \ngal\, is a galaxy selection function that stacks the impacts from all sub-samples:
\begin{equation}
    f = \sum h_k f_k
\end{equation}

We cross-correlate the observed galaxy density field with $f$, which can be viewed as an optimal evaluation of the imaging systematics map:

\begin{multline}
    \langle \delta_{\mathrm{obs}}, f\rangle = \sum_{kk\prime} h_k h_k\langle f_k,f_{k\prime}\rangle \\ +
    \cancel{\sum_{kk\prime} h_{k\prime} \langle \delta_{\mathrm{truth}} f_k, f_{k\prime}\rangle}
    \label{equ:obs-cross-f}
\end{multline}

In Equation \ref{equ:one-term-xi}, the 1st term is the true galaxy clustering signal. The 2nd term is the same as Equation \ref{equ:obs-cross-f}. This term is how people commonly perceive imaging systematics. It is widely studied in various imaging systematics tests and mitigation across all galaxy surveys. The function $f$, which transforms the imaging systematics coordinate to \ngal\, variation, is measured with a wide range of linear and non-linear techniques in imaging systematics studies \cite{ho2012clustering,ross2011ameliorating,elvin2018dark,wagoner2021linear,rodriguez2022dark, chaussidon2022angular,2020MNRAS.495.1613R,johnston2021organised,yan2024kids,kong2020removing}. We call this term \regularsys. 

The 3rd term of Equation \ref{equ:one-term-xi} (also the 6th term in \ref{equ:long-version}) does not appear in \ref{equ:obs-cross-f}. This term contains the new \gpcr\, discussed in this work. \gpcr\, can not be seen in the imaging systematics tests that rely on checking the trend of \ngal\, against systematics maps. The nature of \gpcr\, suggests that even if we have an optimal imaging systematics correction that eliminates all the signals in the 2nd term of equation \ref{equ:one-term-xi}, the corrected galaxy correlation function is still different from the truth.

Strictly speaking, the 3rd term derived in Equation \ref{equ:one-term-xi} is a mixture of \gpcr\, and \textit{multiplicative error} \cite{shafer2015multiplicative}. \textit{Multiplicative error} arises from imaging systematics mitigation methods that correct at the level of galaxy correlation functions. In Appendix \ref{apdx:C}, we present a more detailed discussion on the difference between \gpcr\, and  \textit{multiplicative error}. \gpcr\, can not be transformed into the mathematical form of \textit{multiplicative error}, meaning that the methods that correct \textit{multiplicative error} do not automatically correct \gpcr. In addition, \textit{multiplicative error} does not appear in weight-based methods that apply an optimal weight $w_\mathrm{sys}$ to the observed galaxy density field. However, \gpcr\, still exists for weight-based methods, and we present a detailed discussion in Appendix \ref{apdx:B}. We show that if we measure the correlation function of an optimally weighted galaxy sample, 1st-order terms like the 2nd term in Equation \ref{equ:one-term-xi} vanishes, but the 2nd-order term still exists. After correcting the observed density field with $w_\mathrm{sys}$,  the 3rd term derived in Equation \ref{equ:one-term-xi} is purely induced by \gpcr. 


\subsubsection{Enhancement of clustering amplitude at small $\alpha$}

\label{sec:MpTerm}

We take a closer look at the 3rd term in  Equation \ref{equ:one-term-xi}. 

\begin{equation}
     \sum_{kk'} h_kh_{k'}\langle f_k,f_{k'}\rangle \langle \delta_{\mathrm{truth},k}, \delta_{\mathrm{truth},k\prime}\rangle
     \label{equ:final-form}
\end{equation}

This term can be decomposed into the product of two correlation functions (see Appendix \ref{apdx:A} for further details). Furthermore, we suppose that the observed galaxy density field has been corrected with a weight-based method (See Appendix \ref{apdx:B}). After the weighting, the corrected $f_k$, $f_k^\mathrm{corrected}$ shown in Equation \ref{equ:fcorrected}, fully represents the contribution from \gpcr.

For convenience, we denote $f_{k}^\mathrm{corrected}$ as $f_k$. 
$\delta_{\mathrm{truth},k}$ and $f_k$ are both scalar maps  that vary with sky position $\Omega$. If there are no \gpcr, $\langle \delta_{\mathrm{truth},k}, \delta_{\mathrm{truth},k\prime} \rangle$ would be the same for all $k,k'$. Summing over all $h_kh_{k'}\langle f_k,f_{k'}\rangle$ is 0. However, this term is nonzero if we consider \gpcr, especially for galaxy clustering: galaxies with similar apparent profiles (brightness, color, etc) are also similar in $n(z)$ and $b(z)$. For these galaxies, $f_k, f_{k'}$ are similar, leading to a positive $\langle f_k,f_{k'} \rangle$. Meanwhile, $\langle \delta_{\mathrm{truth},k}, \delta_{\mathrm{truth},k\prime}\rangle$ is also large due to their similarity in $n(z)$ and $b(z)$. On the other hand, if the sub-sample $k,k'$ do not have similar apparent profiles, $\langle f_k,f_{k'}\rangle$ could be small/negative and $\langle \delta_{\mathrm{truth},k}, \delta_{\mathrm{truth},k\prime}\rangle$ is small. Overall, $\langle \delta_{\mathrm{truth},k}, \delta_{\mathrm{truth},k\prime}\rangle$  serves as a `weight' on $\langle f_k,f_{k'} \rangle$: it up-weights large/positive $\langle f_k,f_{k'}\rangle$ and down-weights small/negative $\langle f_k,f_{k'}\rangle$, leading to a positive value on Equation \ref{equ:final-form}.



$f_k$ is particularly large for galaxies near the selection boundaries. The selection function of galaxy sub-sample $k$ could generate large variations in $f_k$. For example, if $k$ is detected in one region and undetected in another, $\langle f_k,f_{k'}\rangle$ could be at the scale of $\sim O(1)$. On the other hand, galaxies far from selection boundaries have a small $\langle f_k,f_{k'}\rangle$.

Aside from the physically motivated reasoning above, we can also view this problem mathematically. The structure of this effect is the systematics map multiplied by the true galaxy density field. If we look at one $\ell, m$ mode from the systematics map and another $\ell\prime, m\prime$ mode from the true galaxy density field, the multiplication of these two mode is:
\begin{equation}
    Y_{\ell m} Y_{\ell' m'} = \sum_{\ell'' = |\ell - \ell'|}^{\ell + \ell'}  C_{\ell m \ell' m'}^{\ell'' m + m'} Y_{\ell'' m + m'}
    \label{equ:ylmylm}
\end{equation}

$C_{\ell m \ell' m'}^{\ell'' m + m'}$ is the \textit{Clebsch–Gordan} coefficient. It ranges from $|\ell-\ell\prime|$ to $\ell+\ell\prime$. Such structure suggests that for an observed $\ell$ mode, it can get contaminated signal from all $\ell$s lower than, or similar to itself. At fixed $\ell, \ell'$, the largest amplitude among all $\ell''$ modes tend to occur at $\ell'' = \ell+\ell' $\cite{edmonds1996angular}. When we go to higher $\ell$s, more $\ell$ modes contribute to the contamination signal, thus we see $|C_\ell|$ biased high at large $\ell$. Since positive ones are up-weighted by $\langle \delta_{\mathrm{truth},k}, \delta_{\mathrm{truth},k\prime}\rangle$, the stacked $C_\ell$ for the whole galaxy sample is biased high. 

Equation \ref{equ:ylmylm} suggests that low $\ell$ modes in a systematics map contaminate more modes than high $\ell$ modes. Indeed, the impact of \gpcr\, is pattern dependent. In Appendix \ref{apdx:D}, we provide toy models of \gpcr\, assuming linear imaging systematics. The toy models suggest that the patterns of the systematics maps play an important role in determining the impact of \gpcr. 

\subsubsection{Comparison with \regularsys: additive vs Multiplicative}
\regularsys\, that causes \ngal\, trends can be viewed as `additive' to correlation function: $\delta' \delta' = \delta \delta + ff$. Here we na\"ively denote the observed correlation function $\delta' \delta'$, the true correlation function $\delta \delta$, and the contribution of imaging systematics $ff$. Meanwhile, \gpcr\, can be viewed as `multiplicative': $\delta'\delta' = ff \cdot\delta \, \delta$. 

The true clustering signal is exponentially small on large scales due to the nature of gravity, while $ff$ does not have such a dramatic change with scales, leaving large-scale systematics mainly contributed by \regularsys. The small-scale imaging systematics correction is typically a by-product of the correction of \regularsys\, on large scales. Since the signal-to-noise of the small-scale clustering signal is much higher than the large-scale ones, the small-scale imaging systematics correction uncertainty induced by \regularsys\, is typically considered small. 

On the other hand, \gpcr\, exerts a multiplicative effect on the correlation function. It scales with the true clustering signal so it is relatively small on large scales, potentially making it undetectable at those scales. However, it could cause a significant enough modification to the small-scale clustering signal. Without properly accounting for this effect, one may underestimate the error budget on small-scale clustering. 

\subsection{Anisotropic redshift and galaxy bias variation}
\label{sec:gpcr}

\begin{figure*}[htbp]
  \centering
  \includegraphics[width=\textwidth]{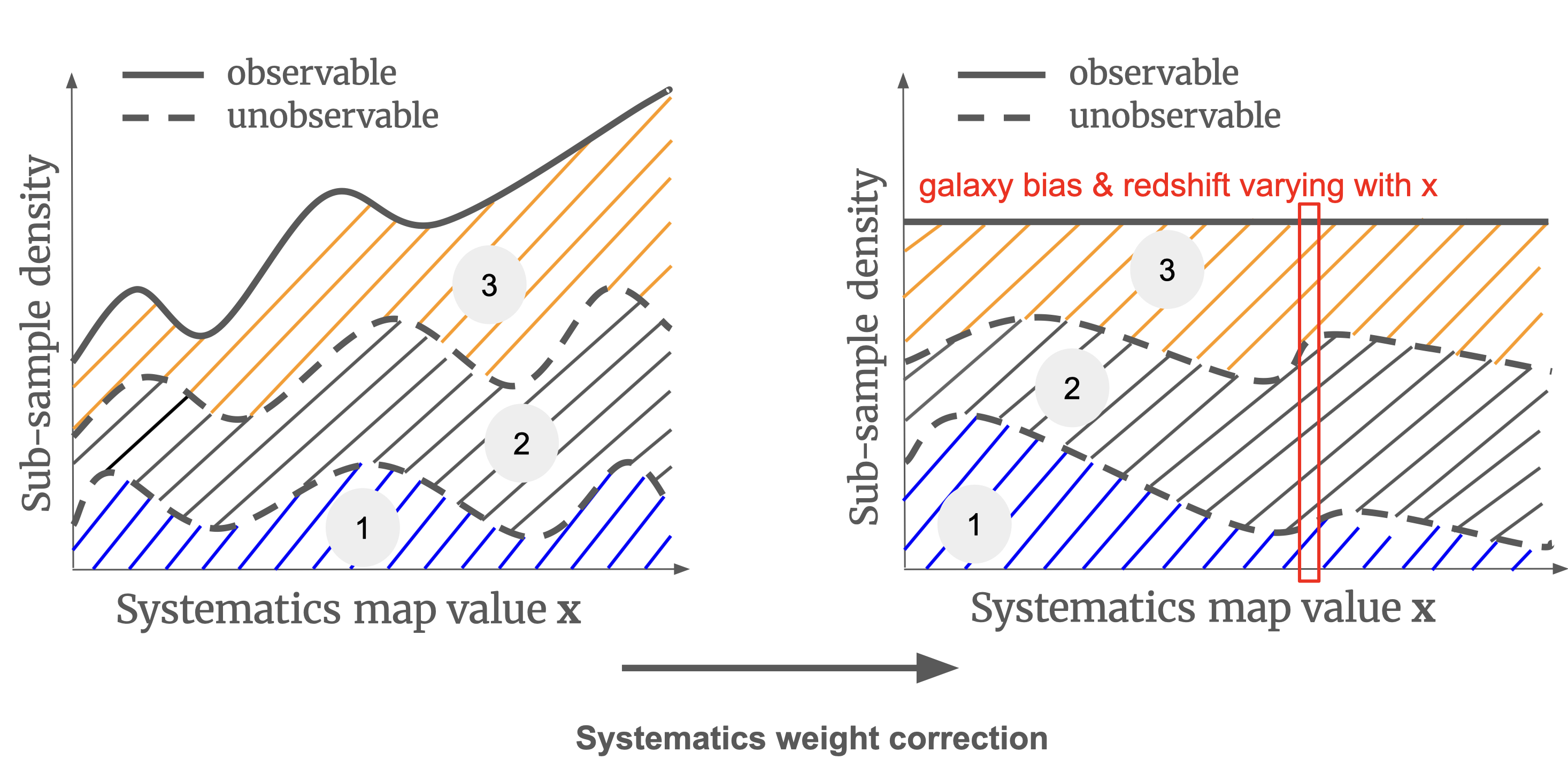}
  \caption{Illustration of how \gpcr\, is formed. A galaxy sample is composed of multiple sub-samples defined by their intrinsic properties like flux, shape, etc. Here we show 3 sub-samples denoted as 1 for blue, 2 for black, and 3 for orange. Looking horizontally, the sub-samples respond differently to the imaging systematics map. The solid black line represents what we can observe: the \ngal\, fluctuation as a function of \textbf{x}. With an optimal imaging systematics correction treating the sample as a whole, this line becomes flat with \textbf{x}. However, the variation in sub-samples still exists. Looking vertically in the right figure, we see that the composition of different sub-samples changes with \textbf{x}. Since each sub-sample has a different redshift and galaxy bias distribution, this leads to a varying redshift and galaxy bias distribution across the entire footprint. }
  \label{fig:sys_correction}
\end{figure*}

We illustrate \gpcr\, in Figure \ref{fig:sys_correction}: when we apply weights to a galaxy sample as a whole, the  overall \ngal\, variation is properly corrected. However, the composition of each sub-sample still changes with the imaging systematics. This is parametrized as $f_k$ in the previous section. 

The galaxy sub-sample $k$ is defined with a large number of parameters like the flux in each band, radius, ellipticity, etc. 
\hui{Ideally, if we split the galaxy sample into multiple sub-samples, and all galaxies each sub-sample $k$ have the same systematics trends $f_k$, then we can treat each sub-sample individually, avoiding the need to model sub-sample systematics.}
However, measuring all $f_k$ accurately is very challenging: if we split the galaxy sample into too many sub-samples, each sample has low density. The true galaxy clustering signal would leak into the measurement of $f_k$. The inability to measure $f_k$ is the fundamental distinction between multiplicative error and \gpcr. We need to find alternative ways to estimate and mitigate \gpcr\, that does not require measuring all $f_k$.  

For the interest of cosmological studies, we care about its impact on the measured galaxy correlation function. For a given galaxy tracer, only two parameters are of interest: the redshift distribution $n(z)$ and the galaxy bias $b(z)$. Thus, we have an alternative way to describe \gpcr:

\textit{The inhomogeneous distribution of galaxies' redshift distribution }$n(z, \textbf{sys})$\textit{ and galaxy bias }$b(z, \textbf{sys})$ \textit{caused by imaging systematics \textbf{sys}. The whole sample's $n(z)$ and $b(z)$ distribution is still accurately measured}.

The two ways of describing \gpcr\, can be seen in Figure \ref{fig:sys_correction}: looking horizontally, we see the density variation of different sub-samples $k$. This variation can be parameterized with $f_k$. Looking vertically, we see the composition of the galaxy sample changing with $\textbf{sys}$, inducing a spatially varying galaxy bias and redshift distribution. This can be parametrized with $n(z, \textbf{sys})$, $b(z, \textbf{sys})$. 

Since we only care about measuring accurate correlation functions, parameterization with $n(z, \textbf{sys})$, $b(z, \textbf{sys})$ is much easier to measure with real data. When dealing with real data, correlation functions are measured by counting all the galaxy-galaxy pairs $DD$ at an angular separation $\alpha$. Then this value is normalized with uniform random pairs $RR$ at the same footprint. When consider \gpcr, the galaxy-galaxy pair counts $DD$ at angular scale $\alpha$ can be written as:

\begin{equation}
     \sum_{IJ} D(\textbf{sys}(\Omega_I))D(\textbf{sys}(\Omega_J))\delta( || \Omega_I - \Omega_J || = \alpha) 
     \label{equ:DDIJ}
\end{equation}

Here the sum of I and J loops over the fully observed sky. The redshift and galaxy bias of $D$ is a function of the systematics map value $\textbf{sys}$: 

\begin{equation}
    D (\textbf{sys}(\Omega_I)) = D \left( b(z, \textbf{sys}(\Omega_I)), n(z, \textbf{sys}(\Omega_I)) \right)
\end{equation}

The amplitude of the correlation function for galaxy clustering is proportional to:

\begin{equation}
    b(z, \textbf{sys}(\Omega_I)) n(z, \textbf{sys}(\Omega_I)) b(z, \textbf{sys}(\Omega_J)) n(z, \textbf{sys}(\Omega_J)) 
\end{equation}

\begin{figure}
    \centering
    \includegraphics[width=\linewidth]{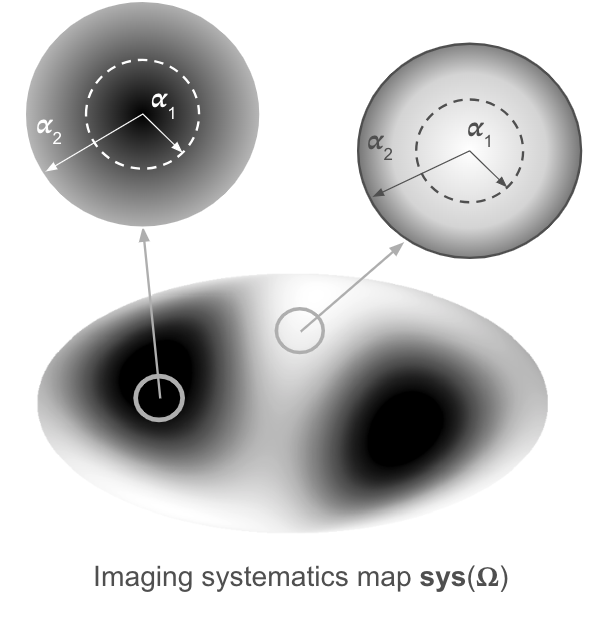}
    \caption{An illustration showing how \gpcr\, modifies the observed galaxy density field and the two-point statistics. We show a fiducial systematics map with its color gradient representing the change in systematics map value $\textbf{sys}(\Omega)$. The redshift and galaxy bias distribution changes with $\textbf{sys}(\Omega)$. The galaxies at sky positions of similar color are more similar to each other. In this example, we see that the color pairs at $\alpha_1$ are more similar to the galaxy pairs at $\alpha_2$. In Section \ref{sec:generalized-framework}, we develop a theoretical framework to describe this phenomenon.}
    \label{fig:lens-source-diff}
\end{figure}

We can interpret Equation \ref{equ:DDIJ} as such: we go over full observed sky with $\Omega_I$. For each $\Omega_I$, we find all galaxies at sky positions $\Omega_J$ that have angular scale $\alpha$ with $\Omega_I$. The galaxy pair counts are reordered: when a galaxy finds its neighbors at a given separation $\alpha$, the expected sample statistics of these neighbors change from sample average $b(z), n(z)$ to the systematics-altered sample statistics ($b(z, \textbf{sys}$), $n(z, \textbf{sys})$). We illustrate this effect in Figure \ref{fig:lens-source-diff}. The galaxy-galaxy pairs at $\alpha_1$ are more similar to each other than the pairs at $\alpha_2$.






\subsubsection{Galaxy correlation function: numerical expression}
\label{sec:generalized-framework}

We assume that the \gpcr\, is the only imaging systematics here. In other words, other systematics that produces \ngal\, trend have been properly corrected. \hui{This is possible with a properly defined imaging systematics weight. We discuss it in more details in Appendix \ref{apdx:B}.}

We split the observed galaxy sample into N sub-samples based on how the imaging systematics shifts the selection function. As long as N is large enough, each sub-sample can be viewed as a clean, homogeneously distributed sample spanning its corresponding footprint. 

We denote the observed tracers for each sub-region $A$ as $D_A$. The region $A$ confined by discretizing the systematics vector $\textbf{sys}$. Different $D_A$ have different sample statistics. Its corresponding random sample is $R_A$. $R_A$ is uniform random within the same footprint as $D_A$. The sum of all $R_{A}$ is a uniform random sample across the whole observed footprint. Under this assumption, the observed galaxy sample is:
\begin{equation}
    D_{\mathrm{obs}} = \sum_{\textbf{sys} = A}^{\textbf{sys} = N} D_A
    \label{equ:d_obs}
\end{equation}

We abbreviate $\sum_I$ $D(\textbf{sys}(\Omega_I)=A)$ as $D_A$. The corresponding randoms are:
\begin{equation}
    R_{\mathrm{tot}} = \sum_{\textbf{sys}=A}^{\textbf{sys}=N} R_A
    \label{equ:r-tot}
\end{equation}

\hui{To elaborate, $D_A, R_A$ can be obtained in the following steps: First, we find the the footprint with imaging systematics value $\textbf{sys}=A$. Next, we populate uniform randoms in this footprint and define the randoms as $R_A$. $D_A$ represents all the galaxies within the footprint with $\textbf{sys}=A$. }$R_A$ is a uniform random sample in the same footprint. The correlation function of a galaxy sample is estimated with the Landy-Szalay \cite{landy1993bias} estimator: 

\begin{equation}
    w_{\mathrm{obs}} = \frac{D_{\mathrm{obs}}D_{\mathrm{obs}} - 2D_{\mathrm{obs}}R_{\mathrm{tot}} + R_{\mathrm{tot}}R_{\mathrm{tot}}}{R_{\mathrm{tot}}R_{\mathrm{tot}}}
\end{equation}

Replacing $D_{\mathrm{obs}}$ and $R_{\mathrm{tot}}$ with all the galaxy sub-samples with equation \ref{equ:d_obs} and \ref{equ:r-tot}:

\begin{equation}
    w_{\mathrm{obs}} = \frac{\sum_{AB}^N (D_A D_B - D_A R_B - R_AD_B + R_A R_B)}{R_{\mathrm{tot}}R_{\mathrm{tot}}}
\end{equation}

We add $R_A R_B$ in the denominator to construct correlation function for each sub-sample: 

\begin{equation}
    w_{obs} = \sum_{ij}^{N} \frac{R_A R_B}{R_{tot}R_{tot}} \frac{D_AD_B - D_AR_B - R_A D_B + R_A R_B}{R_A R_B}
    \label{equ:ls-gpcr}
\end{equation}

Since each sub-sample is a clean and homogeneous sample, $(D_A D_B - D_A R_B - R_A D_B + R_A R_B)/R_A R_B$ is simply the true signal from cosmology, we denote it as $w_{AB}$. We define: 
\begin{equation}
    \mathrm{Win}_{AB} = \frac{R_A R_B}{R_{\mathrm{tot}}R_{\mathrm{tot}}}
    \label{equ:win_ij}
\end{equation}

$\mathrm{Win}_{AB}$ is the cross-correlation of the un-normalized window function for patch $A$ and $B$. It serves as a `hypothetical window function' that modifies the true correlation function signal as a function of $\alpha$. `Hypothetical' comes from the fact that we do not measure this function for a real galaxy sample. 

Equation \ref{equ:ls-gpcr} can be written in the form of: 
\begin{equation}
    w_{\mathrm{obs}}(\alpha) = \sum_{AB} \mathrm{Win}_{AB}(\alpha) \cdot w_{AB}(\alpha)
    \label{equ:generalized-wtheta}
\end{equation}

$w_{AB}(\alpha)$ is the angular cross-correlation function of a galaxy sample with redshift $n(z, \textbf{sys}(\Omega)=A)$ and galaxy bias  $b(z, \textbf{sys}(\Omega)=A)$ and another galaxy sample with $n(z, \textbf{sys}(\Omega)=B)$, $b(z, \textbf{sys}(\Omega)=B)$. The toy model in Section \ref{sec:example-case-ebv} is based on Equation \ref{equ:generalized-wtheta}.


\subsubsection{Galaxy correlation function: analytical expression}
\label{sec:analytical-2nd-order-sys}

We can write the angular correlation function as: 

\begin{multline}
    w(\alpha)_{\mathrm{corrected}} = \\
    \sum_\ell P_\ell(\cos \, \alpha)\int_{||\Omega_1-\Omega_2||=\alpha} dz\, d\Omega_1 d\Omega_2 M_\ell(z) 
    \\ \cdot  n(z, \textbf{sys}(\Omega_1)) n(z, \textbf{sys}(\Omega_2)) \\ \cdot
    b(z, \textbf{sys}(\Omega_1))b(z, \textbf{sys}(\Omega_2))/N(\alpha)  
\end{multline}

Here $\Omega_1, \Omega_2$ are sky angular coordinate ($\theta, \phi$). $M_\ell(z)$ is a complicated function containing equations like the power spectrum P(k), Hubble parameter H(z), integrations over $k$, etc. It depends on the underlying cosmology, but it is independent of the properties of galaxy tracers. $N(\alpha)$ is an normalization factor:
\begin{equation}
    N(\alpha) = \int_{||\Omega_1-\Omega_2||=\alpha} d\Omega_1 d\Omega_2
\end{equation}

We define a parameter that quantifies \gpcr:

\begin{equation}
    \Delta K(z, \textbf{sys}(\Omega)) = \frac{n(z, \textbf{sys}(\Omega))b(z, \textbf{sys}(\Omega))}{n(z)b(z)} - 1
    \label{equ:Kz}
\end{equation}

This is effectively cross-correlating $\Delta K(z, \Omega_{1(2)})$ with a constant field. If the $\Delta K$ map does not contain $\ell=0$ mode, then
\begin{equation}
   \int_{||\Omega_1-\Omega_2||=\alpha} dz\, d\Omega_1 d\Omega_2 \cdot \Delta K\left( z, \textbf{sys}(\Omega_{1(2)})\right)/N(\alpha) = 0
\end{equation}

$\Omega_{1(2)}$ means $\Omega_{1}$ or $\Omega_{2}$. This is effectively cross-correlating $\Delta K(z, \Omega_{1(2)})$ with a constant field. $\ell=0$ corresponds to the mean of $\Delta K$. If it is non-zero, the result is a function independent of $\alpha$

\begin{multline}
   \int_{||\Omega_1-\Omega_2||=\alpha} dz\, d\Omega_1 d\Omega_2 \\ \cdot \Delta K(z, \textbf{sys}(\Omega_{1(2)}))/N(\alpha) = \Delta \bar{K}(z)
\end{multline}

Since we already assume that the mean of $n(z, \Omega$) is $n(z)$, and the mean of $b(z, \Omega)$ is $b(z)$, $\Delta \bar{K}(z)$ is small and likely negligible. $w(\alpha)_{\mathrm{corrected}}$ can be re-written as:

\begin{multline}
    w(\alpha)_{\mathrm{corrected}} = w(\alpha)_{\mathrm{truth}} + \sum_{\ell} P_{\ell}(\cos \,\alpha)\\
     \int_{||\Omega_1-\Omega_2||=\alpha} dz\, d\Omega_1 d\Omega_2 M_\ell(z) \cdot n(z)^2b(z)^2  \cdot \\
    \left( 2\Delta \bar{K}(z) + \Delta K(z, \textbf{sys}(\Omega_1)) \cdot \Delta K(z, \textbf{sys}(\Omega_2))/N(\alpha) \right)
\end{multline}

At fixed $z$, the $\theta, \phi$ terms are the correlation function of $\Delta K(z)$: 

\begin{multline}
    w(\alpha)_{\mathrm{corrected}} = w(\alpha)_{\mathrm{truth}} + \\
    \sum_{\ell} P_{\ell}(cos\,\alpha) \int dz M_\ell(z) \cdot n(z)^2b(z)^2  \cdot \\
    \left( 2\Delta \bar{K}(z) + \langle \Delta K, \Delta K\rangle (z, \alpha)\right)
    \label{equ:exact-solution}
\end{multline}

The second term shows that correlation function of $\Delta K$ serves as a `weight' inside the integral that computes the angular correlation function. Since the part related to underlying cosmology does not have a large variation, we can have a na\"ive estimate on the amplitude contributed by \gpcr\, as:

\begin{equation}
    A(\alpha)\sim\frac{ \int dz\, n(z)^2 b(z)^2 \left(  \langle \Delta K, \Delta K \rangle (z,\alpha)+ 2\Delta \bar{K}(z)\right)}{\int dz\, n(z)^2 b(z)^2 }
    \label{equ:sys-amp}
\end{equation}

Practically, we can produce $\Delta K$ maps for a set of redshift slices by estimating their redshift and galaxy bias variation as a function of imaging systematics. Then we use Equation \ref{equ:sys-amp} to have an order-of-magnitude estimate on \gpcr. 

As discussed in Section \ref{sec:MpTerm}, \gpcr\, leads to an overestimation of the galaxy clustering correlation function amplitude. The same conclusion can be drawn more intuitively by modeling it with $\Delta K$, which reflects the anisotropic redshift and galaxy bias distribution. In Figure \ref{fig:lens-source-diff}, the systematics maps causing biased flux typically have a continuous, large-scale pattern. When a galaxy finds its corresponding pair, it tends to find galaxies similar to itself. At an angular scale $\alpha$ of cosmological interest ($< 5$ degrees), the two $\Delta K$ values at angular separation $\alpha$ tend to have the same signs (both positive or both negative). As a result, $\langle \Delta K, \Delta K \rangle$ tend to be positive. Consequently, the amplitude of the galaxy clustering correlation function is biased high. This effect is scale-dependent: it becomes more significant at smaller $\alpha$. Due to the patterned structure in the systematics map, galaxy pairs exhibit greater similarity on smaller scales. Its impact on cosmological inference, especially on neutrino mass estimation, is discussed in Section \ref{sec:clumping-cosmology}.







Meanwhile, the galaxy-galaxy lensing (GGL) and the cosmic shear (CS) amplitude do not have a significant change with \gpcr. Their clustering amplitude does not increase dramatically when changing the galaxy-galaxy pair ordering.  We demonstrate with toy models in Section \ref{sec:ggl} and Section \ref{sec:shear}.
\section{\texorpdfstring{3 $\times$ 2pt statistics of \gpcr\, with a toy model}{3 x 2pt statistics of 2nd order systematics with a toy model}}

\label{sec:example-case-ebv}

\hui{Overall, we expect \gpcr\, to be a small effect. Combined with cosmic variance, it is hard to see this effect by simply looking at the measured correlation function. Thus, we build a toy model with \textbf{unrealistically large} $n(z), b(z)$ variation. This would amplify the impact of \gpcr, making it noticeable. }This toy model helps us understand that the impact of \gpcr\, is large on galaxy clustering, and negligible for GGL and CS. Such configuration ensures that the difference between galaxy clustering and GGL/CS can be visually seen. Despite being an unrealistic toy model, the methodologies presented in this section can be used to estimate \gpcr\, realistically with an actual galaxy sample. 

In this toy model, we use two dust extinction maps. We suppose that the lens and source galaxies are selected through a set of extinction-corrected color cuts. The $E(B-V)$ map used for target selection is the SFD98 $E(B-V)$ map \cite{schlegel1998maps}. Meanwhile, we assume that true dust extinction map is the DESI $E(B-V)$ map \cite{zhou2024stellar}. Though this is a na\"ive assumption,  this map does show improvements in target selection for DESI ELGs \cite{raichoor2023target} compared to the SFD98 map. \huip{As shown in Figure \ref{fig:sys_correction}, the impact of sub-sample systematics can be captured by allowing the galaxy bias b(z), and redshift distribution n(z) to vary as functions of the measured systematics values.} The difference $\Delta E(B-V)$ between the two maps modifies the $n(z), b(z)$ distribution of our galaxy sample. 

\noindent For lens galaxies:
\begin{equation}
    n_l(z, \textbf{sys}) = \mathrm{Gauss}( 0.8 + 5\Delta E(B-V), 0.15)
    \label{equ:III1}
\end{equation}
\begin{equation}
    b_l(z, \textbf{sys}) = 1.5 + 10 \Delta E(B-V)
    \label{equ:III2}
\end{equation}

\noindent For source galaxies: 
\begin{equation}
    n_s(z, \textbf{sys}) = \mathrm{Gauss}( 1.3 + 5\Delta E(B-V), 0.15)
    \label{equ:III3}
\end{equation}

The `Gauss(A,B)' represents a Gaussian distribution with mean A and standard deviation B.  

Both $E(B-V)$ maps are saved in \textsc{healpix} \cite{gorski2005healpix} format with a resolution of 256. We exclude pixels with a difference over 0.05 mag between these two maps. This comprises 0.72\% of the total overlapping region. The 68\%(95\%) range of the $\Delta E(B-V)$ between the two final maps is 0.024(0.052) mag. We split $\Delta E(B-V)$ from -0.05 to +0.05 into 100 equal-size bins. 

We compute the angular (cross-)correlation function $w_{AB}(\alpha)$ with Equation \ref{equ:generalized-wtheta}. $w_{AB}(\alpha)$ is analytically computed using \textsc{pyccl} \cite{chisari2019core}, with redshift and galaxy bias distribution from Equations~(\ref{equ:III1})--(\ref{equ:III3}) as inputs. $w_{AB}(\alpha)$ does not contain cosmic variance. 

We also produce 100$\times$100 (cross-)window functions $\mathrm{Win_{AB}(\alpha)}$ based on the $\Delta E(B-V)$ map: we first produce a \textsc{healpix} grid with a resolution of 4096 (\texttt{res4096}). \texttt{res4096} is served as `randoms' used to compute $\mathrm{Win_{AB}(\theta)}$ from Equation \ref{equ:win_ij}. We attach a $\Delta E(B-V)$ value for each `random' in \texttt{res4096}. For each of the 100 $\Delta E(B-V)$ bins, R$_A$ represents all `randoms' whose attached $\Delta E(B-V)$ is within this bin. We produce 100 `random' maps by filtering out \texttt{res4096} within a corresponding $\Delta E(B-V)$ bin each time. 

We compute angular correlation function with \gpcr\, using $\mathrm{Win_{AB}(\alpha)}$, $w_{AB}(\alpha)$ described above. Since $w_{AB}(\alpha)$ does not contain cosmic variance, the measured angular correlation function is also free of cosmic variance. 

\begin{figure}
    \centering
    \includegraphics[width=\linewidth]{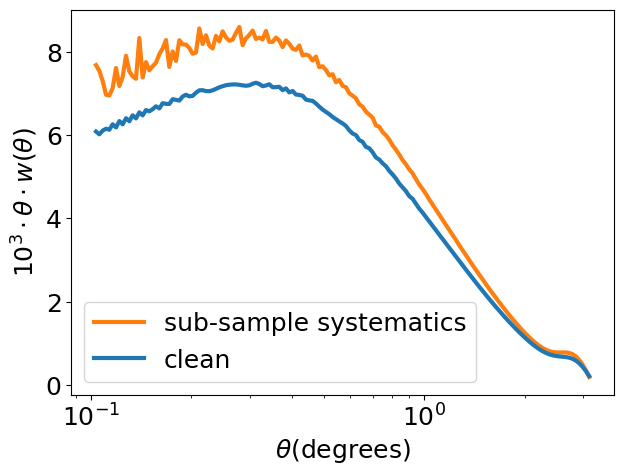}
    \caption{The angular correlation function of the galaxy sample with \gpcr\, (orange) versus the clean sample(blue). The orange curve is derived from equation \ref{equ:generalized-wtheta}. The blue curve is a clean, homogeneous sample theoretically derived with the overall $b(z), n(z)$ identical to the orange curve.}
    \label{fig:ebv-wtheta}
\end{figure}

We first look at the case of galaxy clustering. Figure \ref{fig:ebv-wtheta} shows the difference between the observed angular correlation function $w(\alpha)_{\mathrm{obs}}$ and the angular correlation function generated by the `whole sample' redshift distribution $w(\alpha)_\mathrm{true}$. The wiggles in small angular scales are a result of not having enough randoms in map \texttt{res4096} that computes the window functions $\mathrm{Win}_{AB}(\alpha)$. We also tested a version with 1/8 of current random counts (\texttt{res512}) and found that the small-scale wiggles are larger, but the overall shape of this angular correlation function does not change. The precision meets the requirements for this work. If needed, the precision can be increased by using more randoms to compute all the window functions. As predicted by \gpcr, the amplitude of $w(\alpha)_\mathrm{obs}$ is higher than $w(\alpha)_\mathrm{true}$ on small angular scales. In fact, the $w_\mathrm{obs}(\alpha)$ shown in Figure~\ref{fig:ebv-wtheta} is entirely analytical. The apparent wiggles are not noise; rather, they result from the discontinuous definition of the window function Win$_{AB}(\alpha)$.

When comparing any two $E(B-V)$ maps currently available \cite{schlegel1998maps, lenz2017new, aghanim2016planck, zhou2024stellar, green20193d, mudur2023stellar}, we find that any two of them exhibit distinct patterns\footnote{\href{https://github.com/DriftingPig/dust_map_compare}{A comparison of different dust maps}}. With no knowledge of which map is more accurate, their difference can be viewed as the $E(B-V)$ error. It is vital to have further investigations of $E(B-V)$ to improve its accuracy. 

\begin{figure}
    \begin{subfigure}{\linewidth}
        \includegraphics[width=0.9\linewidth]{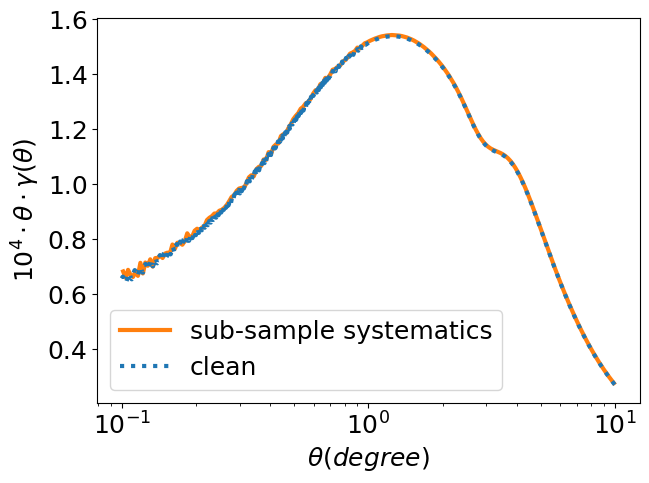}
        \caption{Galaxy Galaxy Lensing tangential shear $\gamma_t$}
    \end{subfigure}

    \begin{subfigure}{\linewidth}
        \includegraphics[width=0.9\linewidth]{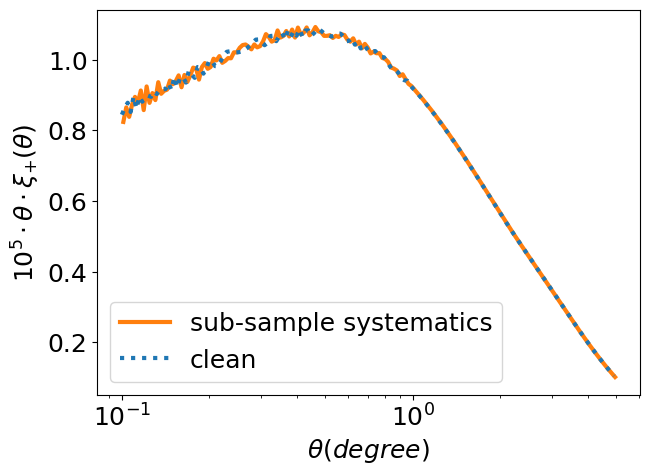}
        \caption{Cosmic Shear correlation function $\xi_+$}
    \end{subfigure}

    \begin{subfigure}{\linewidth}
        \includegraphics[width=0.9\linewidth]{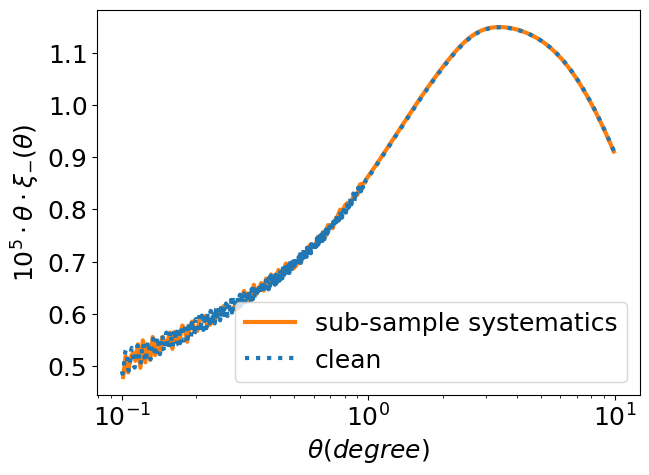}
        \caption{Cosmic Shear correlation function $\xi_-$}
    \end{subfigure}
    \caption{Correlation function for the case of Galaxy Galaxy Lensing (1st row) and Cosmic shear (2nd and 3rd row). Orange curves are the observed correlation function with \gpcr\, and blue curves are the correlation function computed from the $n(z),b(z)$ of the whole sample. No significant difference due to \gpcr\, is detected for all these cases. }
    \label{fig:clumping-lensing}
\end{figure}

Next, we look at the case for GGL and CS. Figure \ref{fig:clumping-lensing} shows the signal measured from GGL and CS. We find no significant bias between the signal measured from the sample with (orange) and without (blue) \gpcr. Section \ref{sec:ggl-negligible}
 and \ref{sec:cs-negligible} analytically estimate the impact of \gpcr\, on GGL and CS, and found that the effect is indeed negligible to 1st order. 

\cite{lizancos2023impact} discusses the impact on GGL and CS to 3rd order. According to their work, the correlation functions of our toy model has a fractional positive bias at the order of $10^{-3}$. However, the uncertainties caused by the overall sample statistics (the mean and shape of $n(z)$), would be much larger than this second order bias.

\section{Analytical analysis of \gpcr\, with toy models }
\label{sec:three-toy-models}

If we draw a horizontal line in Figure \ref{fig:sys_correction} (right), this line could touch the dashed lines. The galaxy sample around this line is changed from one type to another as a function of the systematics map \textbf{sys}.  \gpcrB\, is formed by a collection of such change. Thus, we can investigate a `unit' of \gpcr: we consider a galaxy sample combining two sub-samples. The two types of galaxies in a galaxy sample have different galaxy bias and/or redshift distribution. The whole sample has no \ngal\, trend against the survey property map \textbf{sys}. However, only one sub-sample is seen at a given sky position, and it is determined by map \textbf{sys}. (Each sub-sample scales with a survey property map \textbf{sys} as a step function, and their trend is opposite). For simplicity, we also assume that the two tracers occupy the full sky.

Although the two sub-samples have \ngal\, trend against map \textbf{sys}, this trend is undetectable. We observe the two sub-samples as a whole. The observed sample, a combination of the two types of galaxies, does not exhibit a \ngal\, trend with this map \textbf{sys}. Imaging systematics mitigation methods that rely on probing \ngal\, trend against \textbf{sys} can not detect signals from this toy model.

In section \ref{sec:shift_bias}, the two samples are correlated and share the same underlying matter density field. In section \ref{sec:shift_in_z}, the two samples are uncorrelated, meaning they are in different redshift ranges. Section \ref{sec:ggl} considers the problem in the context of galaxy-galaxy lensing (GGL), and Section \ref{sec:shear} considers the problem in the context of cosmic shear (CS).

\subsection{redshift \& galaxy bias shifts in Galaxy Clustering}
\subsubsection{Shift in galaxy bias}
\label{sec:shift_bias}

We assume two galaxy tracers are in the same redshift range, but they have different galaxy bias $b_{1}$, $b_{2}$.  For simplicity, we assume that the matter power spectrum of both tracers are proportional to the same power spectrum $P(k)$: $b_1^2 P(k)$ for tracer 1 and $b_2^2P(k)$ for tracer 2. 

If the two galaxy tracers are homogeneously distributed, the observed density field is:

\begin{equation}
    \delta^{obs}_{tracers,v1} = h_1 b_1 \delta_m + h_2 b_2 \delta_m
\end{equation}

$h_1$, $h_2$ are the fraction of the two tracers in the sample. $\delta_m$ refers to the underlying matter density field. The same redshift distribution ensures that they have the same $\delta_m$. It is equivalent to a sample with a galaxy bias
\begin{equation}
    b_{eff, v1} = h_1 b_1 +h_2 b_2 
    \label{equ:b-shift-bias-v1}
\end{equation}

On the other hand, if they are modified by map \textbf{sys} as a step function, the observed density field is 

\begin{equation}
\delta^{obs}_{tracers,v2} = \begin{cases}
    b_{1} \cdot \delta_{m}, & \textbf{sys} <= 0 \\
    b_{2} \cdot \delta_{m}, & \textbf{sys} > 0
\end{cases}
\label{equ:b-shift-field}
\end{equation}

If we express the observed tracer, which is a sum of the two tracers, in the form of Landy-Szalay \cite{landy1993bias} estimator

\begin{equation}
w = \frac{(D_{1}+D_{2})^2 - 2(D_{1}+D_{2})(R_{1}+R_{2}) + (R_{1}+R_{2})^2}{(R_{1}+R_{2})^2}
\label{equ:wtheta}
\end{equation}
$D_{1}$, $D_{2}$ represents the associated tracers and $R_{1}$, $R_{2}$ represents their corresponding randoms.

We define the window function between tracer $i$ and $j$ as: 
\begin{equation}
\mathrm{Win}_{ij}  = \frac{R_{i}R_{j}}{(R_{1}+R_{2})^2}
\end{equation}

The window function has property:

\begin{equation}
    \mathrm{Win}_{11} + \mathrm{Win}_{22} + \mathrm{Win}_{12} + \mathrm{Win}_{21} = 1
\end{equation}

\begin{equation}
    \mathrm{Win}_{12} = \mathrm{Win}_{21}
\end{equation}

\begin{equation}
    \mathrm{Win}_{11} + \mathrm{Win}_{12} = \frac{R_1}{R_1+R_2} = h_1
\end{equation}

\begin{equation}
    \mathrm{Win}_{22} + \mathrm{Win}_{21} = \frac{R_2}{R_1+R_2} = h_2
\end{equation}

We define 
\begin{equation}
    \Delta \mathrm{Win}_1 = 1 - \mathrm{Win}_{11}/h_{1}
    \label{equ:delta-w}
\end{equation}
\begin{equation}
    \Delta \mathrm{Win}_2 = 1 - \mathrm{Win}_{22}/h_{2}
\end{equation}
We have 
\begin{equation}
h_{1} \Delta  \mathrm{Win}_{1} = h_{2} \Delta  \mathrm{Win}_{2} 
\label{equation:delta-w}
\end{equation}

Equation \ref{equ:wtheta} can be expressed as: 

\begin{multline}
w = \left( b_{1}^2 \cdot \mathrm{Win}_{11} + b_{2}^2 \cdot \mathrm{Win}_{22} \right. \\
\left. + b_{1}b_{2}\cdot (1 - \mathrm{Win}_{11} - \mathrm{Win}_{22}) \right) \cdot w_{m}
\label{equ:wm}
\end{multline}
$w_{m}$ is the correlation function for the underlying matter density field. 

The above equation can be re-written as 
\begin{multline}
w = -(h_{1}b_{1} \Delta \mathrm{Win}_{1}-h_{2}b_{2} \Delta \mathrm{Win}_{2})(b_{1}-b_{2}) \cdot w_{m} + \\(h_{1}b_{1}^2+h_{2}b_{2}^2)\cdot w_{m}
\end{multline}
According to Equation \ref{equation:delta-w}, this is
\begin{equation}
w = -h_{1}\Delta \mathrm{Win}_{1}(b_{1}-b_{2})^2 \cdot w_{m} + (h_{1}b_{1}^2+h_{2}b_{2}^2)\cdot w_{m}
\label{equ:shift-bias-final}
\end{equation}

If $h_1\Delta \mathrm{Win}_1$ is small, this function is close to a galaxy sample with galaxy bias 
\begin{equation}
    b_{eff,v2} = (h_1b_1^2+h_2b_2^2)^{1/2}
    \label{equ:b-shift-bias-v2}
\end{equation}

In fact, with this `hypothetical window', the amplitude of the measured power spectrum is between $b_{eff,v1}$ (Equation \ref{equ:b-shift-bias-v1}) and $b_{eff,v2}$, and it depends on the distribution of tracer 1 and 2 in the footprint. We validate the derived equations with a toy mock in Appendix \ref{sec:bshift-mock}. 

\subsubsection{Shift in redshift}
\label{sec:shift_in_z}

Following similar configurations as Section \ref{sec:shift_bias}, we consider the two tracers with different redshift distribution. For simplicity, we assume the two tracers have no overlap in redshift space and thus are un-correlated with each other. If the two tracers are uniformly distributed, the observed density field is
\begin{equation}
    \delta^{\mathrm{obs}}_{\mathrm{tracers}} = h_1 b_1 \delta_1 + h_2 b_2 \delta_2
\end{equation}
Here $\delta_1$, $\delta_2$ are the underlying matter density field for tracer 1 and 2. For simplicity, we assume that the angular correlation function of $\delta_1$, $\delta_2$ are equal. 
\begin{equation}
    \langle \delta_1,\delta_1\rangle = \langle \delta_2, \delta_2\rangle = w_m
    \label{equ:z-shift-wm}
\end{equation}
This is not true for real galaxy samples, but we impose this assumption so that the conclusions in this Section can be compared to Section \ref{sec:shift_bias}, where the tracers are correlated. Under our assumption, the angular correlation function of $\delta^{\mathrm{obs}}_{\mathrm{tracers}}$ is 
\begin{equation}
    w = (h_1^2b_1^2 + h_2^2 b_2^2)w_m
\end{equation}
The effective galaxy bias for this combined sample is:
\begin{equation}
    b_{eff, v3} = (h_1^2b_1^2 + h_2^2 b_2^2)^{1/2}
    \label{equ:beff-v3}
\end{equation}

On the other hand, if the tracers are modified by the survey property map \textbf{sys} in the form of a step function, the observed density field is: 

\begin{equation}
\delta^{\mathrm{obs}}_\mathrm{{tracers}} = \begin{cases}
    b_1\delta_1, & \textbf{sys} <= 0 \\
    b_2\delta_{2}, & \textbf{sys} > 0
\end{cases}
\label{equ:delta_obs_z_shift}
\end{equation}

\noindent We can express the angular correlation function in the same ways as Equation \ref{equ:wtheta} (Landy-Szalay estimator). Since the two sample are uncorrelated, $w_{12}=w_{21}=0$, the angular correlation function is

\begin{equation}
\text{$w = w_{m} \cdot (b_1^2\mathrm{Win}_{11} + b_2^2\mathrm{Win}_{22})$}
\label{equ:w_obs_z_shift}
\end{equation}

A similar derivation as in Section \ref{sec:shift_bias} gives us the observed angular correlation function 

\begin{equation}
    w = -(b_1^2+b_2^2)h_{1}\Delta \mathrm{Win}_1 \cdot w_{m} + (h_1 b_1^2 + h_2 b_2^2)w_{m} 
    \label{equ:w_obs_z_shift_clumping}
\end{equation}

Here $\Delta \mathrm{Win}_1$ is defined in Equation \ref{equ:delta-w}. When $h_1\Delta \mathrm{Win}_1$ is small, the observed power spectrum is close to the power spectrum of a galaxy sample with bias 
\begin{equation}
    b_{eff,v4} = (h_1b_1^2 + h_2b_2^2)^{1/2}
    \label{equ:b_eff_v4}
\end{equation}

Compared with $b_{eff,v3}$ in Equation \ref{equ:beff-v3}, we find that the amplitude of the galaxy power spectrum is higher with \gpcr\, effect. For a more general case, the observed power spectrum has an amplitude between $b_{eff,v3}$ and $b_{eff,v4}$, depending on the footprint that produces $\Delta \mathrm{Win}_1$, and how much galaxies are affected by the \gpcr\, effect.  

\subsubsection{Cross-correlation with a clean galaxy sample}
\label{sec:clean-cross}
For the toy model described in Equation \ref{equ:b-shift-field}, the effective bias measured from auto-correlation of the observed field is close to $b_{eff,v2}$ in Equation \ref{equ:b-shift-bias-v2}. However, the effective bias is different when this `contaminated sample' is cross-correlated with a clean sample. In such a cross-correlation scenario, we can combine $\mathrm{Win}_{i1}$ and $\mathrm{Win}_{i2}$ as $\mathrm{Win}_{i}$, as the second sample has the same property inside/outside the `hypothetical window', and do not distinguish itself between $R_1$ and $R_2$ regions:

\begin{equation}
    \mathrm{Win}_1 = \frac{R_1(R_1+R_2)}{(R_1+R_2)^2} = h_{1}
\end{equation}
\begin{equation}
    \mathrm{Win}_2 = \frac{R_2(R_1+R_2)}{(R_1+R_2)^2} = h_{2}
\end{equation}

As the window functions are constant, there is no signal from this `hypothetical window'. The angular correlation function is simply

\begin{equation}
    w = (h_{1} b_1 +h_{2} b_2) w_m
    \label{equ:cl-gm-window}
\end{equation}
Here the effective bias is 

\begin{equation}
    b_{\mathrm{eff}, \mathrm{clean}} = h_{1} b_1 +h_{2} b_2
\end{equation}

This effective galaxy bias equals to $b_{eff,v1}$ in Equation \ref{equ:b-shift-bias-v1}, meaning that we measure the same result as if the tracers have no systematics. In conclusion, \gpcr\, does not impact the cross-correlation measurement between a clean and a contaminated sample.

\subsection{Redshift \& galaxy bias shifts in galaxy-galaxy lensing (GGL) }
\label{sec:ggl}
In the context of GGL \cite{prat2022dark, mandelbaum2005systematic, lange2024systematic,chang2013effective,blandford1991distortion}, source galaxies \cite{gatti2021dark} are distant galaxies whose profiles are modified by the foreground matter density field. Lens galaxies are the biased tracers of the same matter density field. 

In GGL, the estimator is the tangential shear 

\begin{equation}
    \gamma_t (\theta) \sim \frac{\sum_{ls} e_{t,ls}}{N_{ls,tot}}
    \label{equ:gamma-t}
\end{equation}

$e_{t,ls}$ is the tangential ellipticity of a source galaxy relative to a lens galaxy. $N_{ls,tot}$ is the total number of the lens-source pairs. In this toy model, we ignore the weights that optimizes the lensing signal, and the calibration factor that translates the measured ellipticity to absolute shear, and the randoms $e_{r,ls}$.

Similar to the discussion in Section \ref{sec:clean-cross}, if either the lens or source galaxies is a clean sample, or they do not have correlated `hypothetical window', the signal would be the same as if they are homogeneously distributed. If they do have correlated `hypothetical window', both lens and sources galaxies are modified by map \textbf{X} in the same way. In this occasion, based on Equation \ref{equ:generalized-wtheta}, the observed tangential shear is:

\begin{equation}
    \gamma_{obs} = \sum_{ij=1}^{2} \mathrm{Win}_{ij}\gamma_{ij}
    \label{equ:gpcr-gammat}
\end{equation}

However, even if this is true, the impact of \gpcr\, on GGL is still small. We discuss the details in the next section.

\subsubsection{Impact Estimation: negligible}
\label{sec:ggl-negligible}
Due to the structure of GGL, impact of \gpcr\, on GGL is in fact negligible. 

The power spectrum of GGL is \cite{kaiser1992weak, lee2022galaxy} 

\begin{equation}
    C_{g\kappa}(\ell) = \int_0^{\infty} \mathrm{d}\chi \, \frac{q_g(k, \chi) q_{\kappa}(\chi)}{\chi^2} P_{g\delta}\left( \frac{\ell+1/2}{\chi}, z(\chi)\right)
    \label{equ:ggl}
\end{equation}

$q_g(k, \chi)$ and $q_{\kappa}(\chi)$ are the density kernel and the lensing kernel. We ignore intrinsic alignment \cite{troxel2015intrinsic} and magnification \cite{elvin2023dark} as these effects are already small. 

\begin{equation}
    q_{\kappa}(\chi) = \frac{3 H_0^2 \Omega_m}{2 c^2} \frac{\chi}{a(\chi)} \int_{\chi}^{\infty} \mathrm{d}\chi' \, \frac{n_{\kappa}(z(\chi')) \, \mathrm{d}z / \mathrm{d}\chi'}{\bar{n}_{\kappa}} \frac{\chi' - \chi}{\chi'}
\end{equation}

\begin{equation}
    q_g(k, \chi) = \frac{n_g(z(\chi))}{\bar{n}_g} \frac{\mathrm{d}z}{\mathrm{d}\chi}
\end{equation}

$\chi$ is the comoving distance of the lens galaxies and $\chi^{\prime}$ is the comoving distance of the source galaxies. Considering the senario where we have only one lens at distance $\chi$ and one source at distance $\chi^{\prime}$. $n_1(z)$ and $n_2(z)$ are $\delta$ function. Equation \ref{equ:ggl} becomes

\begin{equation}
    C_{g\kappa}(\ell) = \frac{3 H_0^2 \Omega_m}{2 c^2} \frac{\chi(\chi^{\prime}-\chi)}{\chi^{\prime}} \frac{1}{\chi^2} P_{g\delta}\left(\frac{\ell+1/2}{\chi}, z(\chi)\right)/a(\chi)
\end{equation}

If we think about measuring the lensing signal as an average signal of all lens-source pairs:
\begin{equation}
    C_{g\kappa}(\ell) \sim \sum_{ij} \frac{\chi_{i}(\chi_{j}^{\prime} -\chi_i)}{\chi_
    {j}^{\prime}} \frac{1}{\chi_i^2} P_i
\end{equation}

Here $a(\chi)$ is merged into $P_i$. In the presence of \gpcr, the lens and source galaxy distribution are shifted. We split the footprint into multiple parts, making each part a homogeneous galaxy sample. The observed lensing signal is the average signal of the full footprint (summing over t) 

\begin{equation}
    C_{g\kappa}(\ell) \sim \sum_{t}\sum_{ij} \frac{\chi_i^t(\chi_{j}^{\prime,t}-\chi_i^t)}{\chi_
    {j}^{\prime,t}} \frac{1}{ (\chi_i^t)^2 } P_i^t
    \label{equ:simply-ggl}
\end{equation}

As we assume that the only systematics is \gpcr, there is no density fluctuation at different footprint. For each lens $i$ and source $j$, we can find its counterpart in different footprint with shifted distance:

\begin{equation}
    \chi_i^t = \chi_i^{0} + \Delta l_i^t
\end{equation}

\begin{equation}
    \chi_{j}^{\prime,t} = \chi_j^{\prime,0} + \Delta s_i^t
\end{equation}

\begin{equation}
    P_i^t = P_i^0 + \Delta p^t
    \label{equ:82}
\end{equation}

The overall bias/redshift distribution for both lens and source galaxies is accurate. Thus, for each galaxy sub-sample, we can hypothetically assign an accurate bias/redshift distribution, making the bias/redshift distribution for each sub-sample accurate. This is a conceptual idea that makes the following derivations easier. Under this assumption, we have:

\begin{equation}
\sum_t \Delta l_i^t = 0
\label{equ:vanish1}
\end{equation}

\begin{equation}
\sum_t \Delta s_j^t = 0
\label{equ:vanish2}
\end{equation}

\begin{equation}
\sum_t \Delta p_i^t = 0
\label{equ:vanish3}
\end{equation}

Note that Equation \ref{equ:82} is valid because each $P_i$ is determined purely by the lens galaxy associated to it. Since we expect that the overall population of the lens galaxies do not change, this term is also valid. The shifting is small compared to the actual distance or power spectrum, so we can Taylor expand Equation \ref{equ:simply-ggl}:


\begin{multline}
    C_{g\kappa}(\ell) \sim \sum_{ij} \frac{\chi_i^0(\chi_{j}^{\prime,0} - \chi_i^0)}{\chi_
    {j}^{\prime,0}} P_i^0 \cdot \left( 1 
    - \sum_t \frac{\Delta l_i^t}{\chi_i^0} \right. \\
    \left. - \sum_t \frac{\Delta s_j^t}{\chi_j^{\prime,0}}
    + \sum_t \frac{(\Delta l_i^t-\Delta s_j^t)}{(\chi^0_i - \chi^{\prime,0}_j)}
    \right. \\
    \left. + \sum_t \frac{\Delta p^t}{P_i^0}
    + O(\mathrm{variable}^2)
    \right)
    \label{equ:simply-ggl-taylor}
\end{multline}

Here `variable' refers to one of the four terms: $\Delta l_i^t/\chi_i^0$, $\Delta s_i^t/\chi_j^{'0}$, $(\Delta l_i^t-\Delta s_i^t)/(\chi^0_i-\chi_j^{'0})$, $\Delta p^t/P^0_i$. We define the true correlation as 1st order. The variable$^2$ term is two orders of magnitude smaller than the true correlation function, so the $O$(variable$^2$) term is at 3rd order. 

Equations \ref{equ:vanish1}--\ref{equ:vanish3} ensures that all first order terms vanish. The second order terms scale with redshift error due to shifting $\Delta z^2$: a 5\% redshift error only contributes to 0.25\% difference, so the effect is negligible. Note this approximation is only valid when the source galaxies are far from lens galaxies. Otherwise, whether or not source galaxies are behind the lens galaxies would have a large impact. We ignore this scenario as the signal would be dominated by other systematics like intrinsic alignment \cite{vlah2020eft, samuroff2021advances, krause2016impact,prat2022dark}.

In all, the impact of \gpcr\, on GGL measurement is much smaller than galaxy clustering. In the next section, we show the same conclusion also applies to Cosmic Shear.

\subsection{Redshift shift in Cosmic Shear (CS)}
\label{sec:shear}
The estimator \cite{secco2022dark} in CS \cite{asgari2021kids} is the relative pointing of two galaxies at separation $\theta$. 
\begin{equation}
   \xi_{\pm} (\theta) \sim \frac{\sum_{a,b} e_{t,a}e_{t,b} \pm e_{\times,a}e_{\times,b}}{N_{tot}}
\end{equation}
$e_{t}$ and $e_{\times}$ are the tangential and cross components of the ellipticity. This equation follows the same structure as Equation \ref{equ:gamma-t}, so similar derivations can also be applied to this equation. Eventually, we can reach a similar conclusion as in GGL: 

\begin{itemize}
    \item When one sample is contaminated by map \textbf{X} and the other is not, \gpcr\, does not impact the measured correlation function. 
    \item When both samples are contaminated by map \textbf{X}, the observed correlation function is altered (similar structure as Equation \ref{equ:gpcr-gammat}): 
\end{itemize}

\begin{equation}
        \xi_{\pm} = \sum_{ij=1}^2 \mathrm{Win}_{ij} \xi_\pm^{ij}
\end{equation}

However, the impact of \gpcr\, on Cosmic shear is still negligible. We present analysis in the following section. 

\subsubsection{Impact Estimation: negligible}
\label{sec:cs-negligible}
The power spectrum of Cosmic shear measurement is \cite{secco2022dark, loverde2008extended}

\begin{equation}
    C^{ij}_\ell = \int_0^{\infty} d\chi \frac{q^i(\chi)q^j(\chi)}{\chi^2} P_\delta\left(\frac{\ell + 1/2}{\chi}, z(\chi)\right)
    \label{equ:cosmic-shear-cl}
\end{equation}
$q^{i}(\chi)$ is the lensing efficiency kernel. $i$, $j$ refers to the tomographic bins for the source samples. 

\begin{equation}
    q^i(\chi) = \frac{3H_0^2\Omega_{\text{m}}}{2c^2} \frac{\chi}{a(\chi)} \int_\chi^{\infty} d\chi^{\prime} \, n^i(z(\chi^{\prime})) \frac{dz}{d\chi^{\prime}} \frac{\chi^{\prime}-\chi}{\chi^{\prime}}
\end{equation}

Similarly, if we look at a pair of galaxies $m$, $n$ at comoving distance $\chi^{\prime}_{im}$ and $\chi^{\prime}_{jn}$, and assuming the redshift distribution are $\delta$ function, equation \ref{equ:cosmic-shear-cl} becomes 

\begin{equation}
    C^{ij}_\ell = \left(\frac{3H_0^2\Omega_{\text{m}}}{2c^2}\right)^2 \int_{0}^{\chi_H} d\chi \frac{P_\delta(\chi)}{a(\chi)^2} \frac{\chi^{\prime}_{im} - \chi}{\chi^{\prime}_{im} } \cdot\frac{\chi^{\prime}_{jn} - \chi}{\chi^{\prime}_{jn} } 
    \label{equ:cosmic-shear-pre-clumping}
\end{equation}

\begin{equation}
    \chi_H = \mathrm{min}(\chi^{\prime}_{im}, \chi^{\prime}_{jn})
\end{equation}

Similar to GGL, the total observed signal is a sum of all samples $m$, $n$. If only \gpcr\, exists, the source samples in different footprint $t$ have stable number counts. Hypothetically, we order all source samples from low to high redshift, and have a one-to-one match for sources. At different footprint $t$, the galaxy sample can be slightly shifted as $\Delta s_{mt}$ and $\Delta s'_{nt}$. As we assume the overall distance distribution is accurate, for each sample $m$ or $n$, we can use their average distance as a reference. Using similar arguments in Section \ref{sec:ggl-negligible}, we have:

\begin{equation}
   \sum_t \Delta s_{mt} = 0
\end{equation}

\begin{equation}
   \sum_t \Delta s'_{nt} = 0
\end{equation}

We put these terms in the correlation function:

\begin{multline}
    C^{ij}_\ell \sim \sum_{mnt} \int_0^{\chi_H-\Delta_{mn}} d\chi \frac{P_\delta(\chi)}{a(\chi)^2} \frac{\chi^{\prime}_{im} + \Delta s_{mt} - \chi}{\chi^{\prime}_{im} + \Delta s_{mt}} \\
    \cdot\frac{\chi^{\prime}_{jn} + \Delta s_{nt} - \chi}{\chi^{\prime}_{jn} + \Delta s_{nt}} + \\
    \int_{\chi_H-\Delta_{mn}}^{\chi_H+\Delta^{\prime}_{mnt}} \frac{P_\delta(\chi)}{a(\chi)^2} \frac{\chi^{\prime}_{im} + \Delta s_{mt} - \chi}{\chi^{\prime}_{im} + \Delta s_{mt}} \\
    \cdot\frac{\chi^{\prime}_{jn} + \Delta s_{nt} - \chi}{\chi^{\prime}_{jn} + \Delta s_{nt}}
    \label{equ:cosmic-shear-clumping}
\end{multline}

$\Delta_{mn}$, $\Delta_{mnt}$ appears because when the sample $m$, $n$ are shifted, the integration range changes according to the closest source galaxy. Since $\Delta_{mn}$ is small, and either $\chi^{\prime}_{im} - \chi 
  (+ \Delta s_{mt} )$ or $\chi^{\prime}_{jn} - \chi 
 (+\Delta s^{\prime}_{nt})$ is small when $\chi$ is close to $\chi_H$ (one of the source galaxy). Thus, the terms concerning $\Delta_{mnt}$ are of second order.

We define
\begin{equation}
    F(\chi) = \frac{P_\delta(\chi)}{a(\chi)^2} \frac{\chi^{\prime}_{im} - \chi}{\chi^{\prime}_{im}} \cdot\frac{\chi^{\prime}_{jn}-\chi}{\chi^{\prime}_{jn}}
\end{equation}

Equation \ref{equ:cosmic-shear-clumping} becomes

\begin{multline}
    C^{ij}_\ell \sim \sum_{mn}\int_0^{\chi_H} d\chi F(\chi) [ 1+\sum_t \Delta s_{mt}\frac{\chi}{\chi'_{im}-\chi} + \\
    \sum_t \Delta s_{nt}\frac{\chi}{\chi'_{in}-\chi} + O(\mathrm{variable}^2) ]
    \label{equ:cs-1st-order}
\end{multline}

Here `variable' refers to either $\Delta s_{mt}$ or $\Delta s_{nt}$. All first-order terms in Equation \ref{equ:cs-1st-order} vanish and the impact of \gpcr\, on cosmic shear is negligible at 1st order. 

\subsection{Magnification and Intrinsic Alignment(IA)}
The kernel used to compute the correlation function signal contributed by magnification is in a similar mathematical structure as GGL/CS. Thus, the impact of magnification is also negligible at first order. In addition, magnification is a small effect itself, so we can ignore the \gpcr's impact on magnification. 

The kernel used to compute the correlation function signal contributed by IA is similar to galaxy clustering. Intuitively, we can think of a lens-source sample pair that has more radial overlap in some regions than others. This would produce larger intrinsic alignment signal than if they are homogeneously distributed. However, the impact of IA is small for standard cosmological probes. A percentage change on IA is negligible. Meanwhile, it could matter for new cosmological probes. For example, \cite{okumura2020anisotropies} discusses the use of IA as a cosmological probe, and \cite{cross2024inverse} talks about inverse GGL, where IA is a dominant effect. It is necessary to estimate the impact of \gpcr\, for these new cosmological probes. 

\section{\texorpdfstring{Impact of \gpcr\, on \\ cosmological measurements}{Impact of \gpcr\, on cosmological measurements}}
\label{sec:clumping-cosmology}

\subsection{\texorpdfstring{3 $\times$ 2 pt statistics}{3 x 2 pt statistics}}

The analysis in Section \ref{sec:generalized-framework} and the toy models in Section \ref{sec:three-toy-models} all point to a conclusion: \gpcr\, always makes the observed galaxy clustering correlation function higher. Meanwhile, it does not have a noticeable impact on GGL and CS. 

Such variation is different from the systematics produced by photometric redshift error \cite{ma2006effects} (photo-z error). The photo-z error of the lens galaxy can affect both galaxy clustering and galaxy-galaxy lensing, a different variation mode compared to \gpcr. It is typically parametrized as bias in mean redshift, width of distribution, and photo-z outliers \cite{graham2020photometric, graham2017photometric}. For lens galaxies, all these parametrization cause changes in the galaxy correlation function for both galaxy clustering and galaxy-galaxy lensing. Meanwhile, \gpcr\, only impacts galaxy clustering. Thus, this systematics is not the same as photo-z error, and cannot be fully absorbed into the nuisance parameters describing the photo-z error.

\subsection{Degeneracy with neutrino mass measurement}

Terrestrial experiments \cite{abe2011t2k, acero2019first,himmel2020new} on neutrino oscillation \cite{esteban2020fate}  measure a total neutrino mass $\sum m_{\nu}>$ $0.059$ $\mathrm{eV}$. This value is already close to the upper limit from recent cosmological measurements using DESI \cite{adame2023early} and Planck \cite{aghanim2020planck} CMB data \cite{adame2024desi, adame2024desi7}, which measures $\sum m_{\nu} < $ 0.071 eV at 95\% confidence level with a positive neutrino mass prior and assuming a $\Lambda \mathrm{CDM}$ model. These results leave a narrow window for $\sum m_{\nu}$, so people have started to discuss potential neutrino mass tension between cosmological and terrestrial experiments \cite{di2022neutrino, gariazzo2023quantifying}. Forecast of neutrino mass measurement with LSST and CMB data suggests an uncertainty of 0.07 eV \cite{abell2009lsst}.


As discussed in Section \ref{sec:MpTerm}, \gpcr\, enhances the galaxy clustering amplitude at small scales. The mode is similar to negatively biasing the neutrino mass measurement. 

We use a toy model to demonstrate our argument. We assume that a variation of galaxy bias due to $E(B-V)$ map error:

\begin{equation}
    b(\textbf{sys}) = 1.5 + 15 \cdot \Delta {E(B-V)}
    \label{equ:bias-variation}
\end{equation}

The observed correlation function is:

\begin{equation}
    <\delta_{\mathrm{obs}},  \delta_{\mathrm{obs}}> = <b(\textbf{sys})\delta_{\mathrm{truth}}, b(\textbf{sys})\delta_{\mathrm{truth}}> 
\end{equation}

\hui{We choose 15$\cdot \Delta E(B-V)$ to model the galaxy bias variation. Under this choice, 1$\sigma$ variation in galaxy bias is 0.17, about 13\% of the mean galaxy bias. A study on DES \texttt{redMaGiC} sample introduces an $X_\mathrm{lens}$ factor (Figure 12 in \cite{pandey2022dark}). $X_\mathrm{lens}$ is the ratio of best-fit galaxy bias measured from galaxy-galaxy lensing and galaxy clustering. $X_\mathrm{lens}$ fluctuates in different sky patches. The 1$\sigma$ range is roughly +/-10\%. Sub-sample systematics can produce such $X_\mathrm{lens}$ effect, as it makes the galaxy clustering amplitude higher, while has negligible impact on correlation functions derived from galaxy-galaxy lensing. \huip{A 13\% fluctuation in galaxy bias can produce similar $X_\mathrm{lens}$ effects measured from \texttt{redMaGiC}.} Our analysis choice in Equation \ref{equ:bias-variation} is based on this observation.} Overall, this toy model may have been exaggerating the impact of $\Delta {E(B-V)}$. However, the level of galaxy bias variation is possible for real galaxies if we consider all possible systematics.

\begin{figure}
    \centering
    \includegraphics[width=0.98\linewidth]{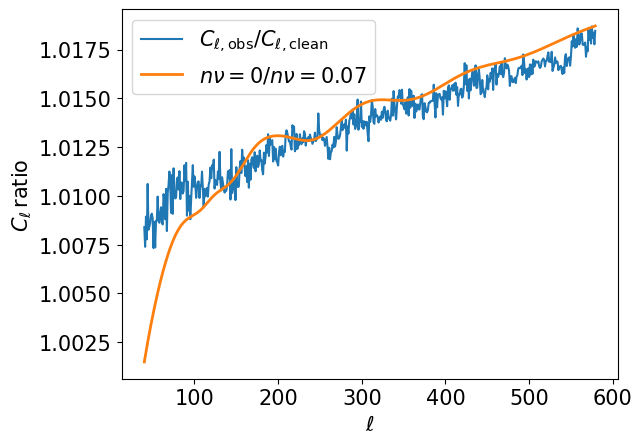}
    \caption{Blue: A galaxy sample with bias variation, divided by a homogeneous sample. The sample has a neutrino mass of 0.07 eV. Orange: Theoretical correlation function of neutrino mass 0 eV, divided by a curve with the same cosmology but having neutrino mass 0.07 eV. }
    \label{fig:neutrino}
\end{figure}

In Figure \ref{fig:neutrino}, we show how the ratio of observed and true $C_\ell$ have a positive `systematics trend'. This `systematics trend' is similar to the trend produced by the ratio of $C_\ell$ with neutrino mass 0 eV and 0.07 eV. The resemblance suggests that, if we do not consider the \gpcr\, effect, the neutrino mass measurement could be negatively biased, producing a `tension' with the results from the terrestrial experiments. 

An actual cosmological inference probes multi-dimensional parameter space, so a solid assessment on the impact of \gpcr\, on neutrino mass measurement requires more robust testing. Moreover, the effect of \gpcr\, could be different for different tomographic bins, which could also help break the degeneracy. However, if we do not estimate the \gpcr\, effect, and put a flat prior of $k,b$ in the form of $C_{\ell,\mathrm{obs}} = (k\ell+b)C_{\ell, \mathrm{true}}$ as a systematics likelihood, we may lose all constraining power on the neutrino mass measurement. In conclusion, it is essential to estimate this effect for future precision cosmology.

\subsection{Scale-dependent galaxy bias}

\gpcrB\, suggests that the galaxy bias could vary with some systematics, making the galaxy bias scale-dependent. Such systematics undermines the measurement of scale-dependent bias due to cosmological variations. Primordial non-Gaussianity (PNG) \cite{rezaie2024local, chaussidon2024constraining, mueller2022primordial,ross2013clustering, alonso2015constraining,riquelme2023primordial, dalal2008imprints} relies on measuring the scale-dependent bias on large scales. 

\begin{figure}
    \centering
    \includegraphics[width=0.98\linewidth]{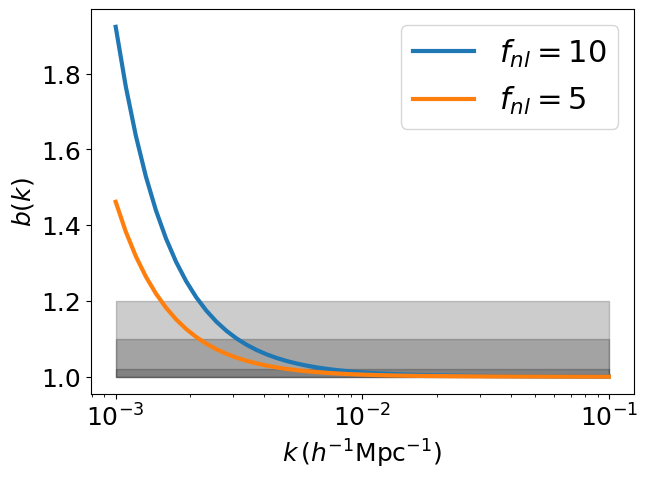}
    \caption{Scale-dependent galaxy bias parametrized by the local primodial non-gaussanity parameter $f_{nl}$. The shaded region indicates 20\%, 10\%, 2\% fluctuation range of galaxy bias fluctiation. }
    \label{fig:PNG}
\end{figure}

Despite this argument, one still need to have a order-of-magnitude estimate to determine whether \gpcr\, is indeed a concern for PNG. Mitigating \regularsys\, is already a challenging task for PNG. Such mitigation often comes with an extra error budget. If \gpcr\, have a much lower impact than this error budget, then it is not necessary to consider this effect. $\Delta K(z)$ from Equation \ref{equ:Kz} is a good starting point. After having an estimate of the level of fluctuation, we can compare it to the amplitude of scale-dependent bias $b(k)$. At large $k$, $b(k)$ scales as $k^{-2}$. Meanwhile, the amplitude change by $\Delta K(z)$ does not have such a dramatic change. Fig \ref{fig:PNG} shows the scale-dependent galaxy bias as a function of wave number $k$ at $f_{nl}=10$ and $f_{nl}=5$. If we can reliably measure $P(k)$ at $k\sim 10^{-3}$, then even a 20\% fluctuation is not a concern. However, if we are unable to correct for \regularsys\, at large scales and have to apply a scale cut, we may enter the regime where \gpcr\, is a concern. In addition, it also depends on the galaxy tracers. For example, the galaxy bias of Quasars (QSOs) is much less sensitive to imaging systematics than the Luminous Red Galaxies (LRGs). From this perspective, PNG measurements for QSOs is less challenging than LRGs \cite{chaussidon2024constraining}. 

In conclusion, it is necessary to assess the error introduced by \gpcr\ and compare it to the errors from \regularsys\, and cosmic variance. If the error from \gpcr\, is comparable in orders-of-magnitude to these other sources, it must be incorporated into the $f_{nl}$ analysis. 

\section{Estimation and Mitigation of \gpcr}
\label{sec:ssi}

It is important to roughly estimate the impact of \gpcr: is it a 0.1\% impact or a 1\% impact on the measurement of the correlation function? This level of estimation helps determine whether it is necessary to incorporate it into our cosmological inference. For example, we might have an error budget of 5\% for a cosmological analysis, but \gpcr\, only contributes 0.1\% of it. In this case, we can argue that \gpcr\, is negligible for our analysis. Fully correcting \gpcr\, is not possible because there are numerous types of galaxy sub-samples. However, we can make sure that the error contributed by \gpcr\, is under control. 

The toy model in Section \ref{sec:shift_in_z} is a good place to start. We can na\"ively suppose that our sample have $h_1$ `outliers' that only appear in some regions of the sky. These `outliers' follow the toy model definition in Section \ref{sec:shift_in_z}. If $h_1=1\%$, $b_1=b_2$, the ratio of the clustering amplitude between the `observed' sample and the `truth' sample is: 

\begin{equation}
    \frac{w_\mathrm{obs}(\alpha)}{w_\mathrm{truth}(\alpha)} = \frac{b_\mathrm{eff, v4}^2}{b_\mathrm{eff, v3}^2} = \frac{1}{h_1^2+h_2^2} \approx 1.02
\end{equation}

The equations for galaxy bias are taken from Equation \ref{equ:beff-v3} and \ref{equ:b_eff_v4}. The clustering amplitude has a 2\% change under this assumption. 

In practice, we need more precise estimation on \gpcr. Moreover, we need to mitigate such variation and control the impact of \gpcr\, below a required error budget. The accuracy of \gpcr\, is determined by the estimation of galaxy bias $b(z, \textbf{sys}(\Omega))$ and redshift distribution $n(z, \textbf{sys}(\Omega))$. Using these variables, we can obtain either an order-of-magnitude estimation with Equation \ref{equ:sys-amp} or a more accurate estimation with Equation \ref{equ:generalized-wtheta}. 

The challenging part is to accurately estimate $n(z, \textbf{sys}(\Omega))$ and $b(z, \textbf{sys}(\Omega))$. In what follows, we discuss the methodology to measure these variables.

\subsection{\hui{Forward modeling the spatially varying redshift and galaxy bias distribution} }

Figure \ref{fig:ssi_diagram} shows the procedure to obtain $n(z, \textbf{sys}(\Omega))$ and $b(z, \textbf{sys}(\Omega))$ from real survey data. In what follows, we explain each step in more detail.

\subsubsection{Preparation of the input catalog}
We start with an input catalog, referred to as the `truth catalog'. This catalog is from a small region with deep imaging. Such regions have a lot more images taken compared to a wide field. Then we assign redshift and galaxy bias to each source in the deep catalog. 

The redshift can be obtained either from a photometric redshift estimate, or, if available, external spectroscopic redshift. 

The galaxy bias, on the other hand, is less straightforward to determine on a per-galaxy basis. One common approach is to divide the sample into luminosity bins, measure the galaxy clustering amplitude for each bin, and model its variation as a function of luminosity. Another promising approach is to use emulators. To implement this, we use galaxies in the wide field to ensure sufficient statistics. For each galaxy, we count all galaxy pairs $DD$ within a certain angular distance range and normalize this number with the corresponding random pairs $RR$ in the sample, computing $DD/RR$ for all galaxies. We then take each galaxy’s flux values in all bands along with its $DD/RR$ value and use these as input to an emulator.
Because the mean $DD/RR$ varies smoothly with source flux, Conditional Normalizing Flows are well-suited for this task: they can learn complex conditional distributions and capture subtle correlations between flux and clustering statistics. After training the CNF emulator, we sample each deep-field galaxy multiple times to obtain a reliable mean $DD/RR$, from which we derive the galaxy bias.

The process of determining redshift and galaxy bias for each source is inevitably subject to some noise. However, our primary goal is to investigate the spatial variation of these quantities, for which a small amount of noise is acceptable.

\begin{figure}[ht]
\centering

\begin{tikzpicture}[
    font=\small,
    box/.style={
        draw,
        rounded corners,
        fill=gray!10,
        align=center,
        minimum width=5.5cm,
        minimum height=2.8cm
    },
    diamondnode/.style={ 
        draw,
        diamond,
        aspect=2,
        align=center,
        inner sep=4pt,
        fill=gray!10
    },
    boxsmall/.style={
        draw,
        rounded corners,
        fill=gray!10,
        align=center,
        minimum width=5cm,
        minimum height=2.4cm
    },
    boxtiny/.style={
        draw,
        rounded corners,
        fill=gray!10,
        align=center,
        minimum width=3.5 cm,
        minimum height=2. cm
    },
    arrow/.style={->, thick},
    node distance=3.5cm and 3.5cm
]

\node[box] (input) {
    \textbf{Each source from input catalog} \\[1mm]
    Input flux in all bands\\
    Input galaxy shape\\
    redshift $z$\\
    galaxy bias $b$\\[2mm]
    \textbf{Systematics map values
}};

\node[diamondnode, below=1.0cm of input] (inj) {
    \textbf{Source}\\
    \textbf{Injection}\\
    \textbf{Emulator}
};

\node[boxsmall, below=1.0cm of inj] (output) {
    \textbf{Each source from output catalog} \\[1mm]
    Output flux in all bands\\
    Output galaxy shapes\\
    redshift $z$\\
    galaxy bias $b$
};

\node[boxtiny, below=1.5cm of output] (select) {
    \textbf{Selected sample} \\[1mm]
    redshift n(z,\textbf{sys}({$\Omega$)})\\[0.2mm]
    galaxy bias b(z,\textbf{sys}({$\Omega$)})
};

\node at (input.north) [above] {\textbf{a HEALPix pixel with sys({$\Omega$)}:}};

\draw[arrow] (input) -- (inj);
\draw[arrow] (inj) -- (output);
\draw[arrow] (output) -- node[right, align=center] {\textbf{Sample} \\ \textbf{selection}} (select);

\end{tikzpicture}

\caption{ A diagram showing the procedure to obtain galaxy bias and redshift distribution in a HEALPix pixel associated with a set of systematics map values.}
\label{fig:ssi_diagram}
\end{figure}

\subsubsection{Building a Source Injection Emulator}

Source Injection refers to the procedure of injecting synthetic galaxies into real images, and processing these modified images in the same way as the original ones. This method is applied to a variety of imaging surveys like \textsc{Balrog} for Dark Energy Survey (DES) \cite{suchyta2016no,everett2022dark,anbajagane2025dark}, \textsc{Obiwan} for Dark Energy Spectroscopic Instrument (DESI) \cite{kong2020removing, kong2024forward,rosado2024mitigating}, and \textsc{SynPipe} for Hyper Suprime-Cam (HSC) \cite{bosch2018hyper}. The Source Injection pipeline for LSST is publicly available \footnote{https://github.com/lsst/source\_injection}. 

Source Injection produces synthetic galaxies that mimic the systematics in real sources. \hui{It automatically includes various systematics effects seen by real galaxies. For example, galaxies could be blended by a nearby source, and have a biased estimate in the flux measurement. The same behavior would also appear in synthetic galaxies from Source Injection.}

Source Injection can partially reproduce the \ngal\, trends induced by the systematics. \hui{The trends that cannot be properly recovered are mainly due to uncertainties in the dust extinction map $E(B-V)$ and zero-point calibration, since source injection does not know the `truth' of these maps. These unrecoverable trends can be mitigated, but it is beyond the scope of this work. The systematics simulated by source injection always exist in real data, but not the other way around. The results obtained from Source Injection can be seen as a lower bound for the \gpcr\, effect. }Here, we assume that source injection can recover all systematic trends. As source injection methods improve, this assumption will become increasingly valid.

Running source injection is a computationally expensive task. For this particular work, we need a large amount of synthetic galaxies in each \textsc{healpix} pixel to obtain a noiseless estimate of $n(z, \textbf{sys}(\Omega))$ and $b(z, \textbf{sys}(\Omega))$. Thus, an emulator for source injection is necessary to efficiently obtain a large amount of synthetic galaxies. Similar to the argument on estimating galaxy bias, Conditional Normalizing Flows (CNFs) provide a powerful framework for building such an emulator. CNFs are capable of learning the conditional distribution of galaxy properties given observational systematics, allowing us to generate a large number of realistic synthetic galaxies in each \textsc{healpix} pixel with minimal computational cost. By training the flow on a subset of explicitly injected sources, the emulator can then rapidly sample new galaxies while preserving the statistical relationships in the original data, making it a practical and scalable solution for our analysis.

\subsubsection{Sample selection and beyond}

Once we have the input catalog and the source injection emulator ready, we produce a large amount of synthetic galaxies in each \textsc{healpix} pixel.Next, we apply a sample selection function, typically consisting of a set of color cuts, to the output synthetic galaxies, and obtain the redshift and galaxy bias distribution of the selected sample. We repeat the same procedure for all \textsc{healpix} pixels. Eventually, we obtain the spatially varying redshift and galaxy bias distribution of this sample. 

The end product can be used to estimate \gpcr\, on the correlation function level. 2-point angular correlation functions can be estimated with Equation \ref{equ:generalized-wtheta}, following the procedure discussed in Section \ref{sec:example-case-ebv}. Alternatively, code provided by \cite{lizancos2023impact} in 
$\ell$-space can be used to perform a similar analysis. For three-dimensional correlation functions (e.g., using spectroscopic redshifts) or higher-order statistics, modeling efforts are not yet available, making this an important avenue for future work.

\section{Conclusion}
\label{sec:conclusion}

We discussed \gpcr: how a spatially varying galaxy bias and redshift distribution subtly change the observed correlation function. Unlike the more familiar 1st order imaging systematics, this effect does not produce a large, spurious signal on large scales. The amplitude of \gpcr\, is proportional to the observed angular correlation function, making it more significant on angular scales with large correlation function amplitude. In contrast, the amplitude of the \regularsys\, varies slowly across all angular scales. Since \gpcr\, do not show alarming features on large scales, they are often ignored in current cosmological analysis.


We introduce the theory behind \gpcr\, in Section \ref{sec:theory}. Different galaxy types have different systematics trend. When correcting them as one sample, the corrected field has a variation in the composition of galaxy types. We demonstrate that \gpcr\, can not be detected by cross-correlating the imaging systematics map with the observed galaxy density field. In other words, it does not show up as a trend in the plot of relative galaxy density \ngal\, versus systematics map value. \gpcr\, always makes the observed galaxy clustering signal higher, and it scales up when going to smaller scales. 


We define a parameter $\Delta K(z, \Omega)$ (Equation \ref{equ:Kz}) which quantifies how $b(z), n(z)$ varies with systematics. $\Delta K(z, \Omega)$ can be used to estimate the impact of \gpcr\, for any specific galaxy tracer. 

Section \ref{sec:example-case-ebv} presents the methodology to estimate \gpcr\, with a realistic galaxy sample. By measuring the hypothetical window functions defined in \ref{equ:win_ij}, we derive the observed correlation function. We found that \gpcr\, is only a concern for galaxy clustering, and it is more significant on small scales.

Section \ref{sec:three-toy-models} shows several toy models when two sub-samples have either different redshift or different galaxy bias. We discussed the situations for Galaxy Clustering, Galaxy-Galaxy Lensing (GGL), and Cosmic Shear (CS). We find that for GGL and CS, terms of \gpcr\, cancels out until 2nd order. Meanwhile, the 2nd order terms galaxy clustering does not cancel out. This explains why \gpcr\, is more significant for galaxy clustering. 


Section \ref{sec:clumping-cosmology} discusses impacts of the \gpcr\, on cosmological measurements. \gpcr\, can mimic the impact of massive neutrinos in the angular power spectra, which could, in principle,  biase our results towards $m_{\nu} = 0$, as shown in Figure \ref{fig:neutrino}. This is particularly relevant in the context of DESI + CMB that have reported $m_{\nu} < 0.07$, almost in tension with direct measurement of neutrino masses \cite{adame2025desi}. For $3\times 2PT$ analysis, \gpcr\, cause an inconsistent measurement of galaxy bias between galaxy-galaxy lensing and galaxy clustering. This is similar to the mysterious $X_\mathrm{lens}$ factor found in the DES \texttt{redMaGiC} studies \cite{pandey2022dark}. It could bias $S_8$, a cosmological parameter that measures the clustering strength of matter in the universe. $S_8$ is sensitive to the amplitude of the measured galaxy correlation function. \gpcr\, produces scale-dependent bias similar to Primordial Non-gaussianity (PNG). The systematics of PNG could be dominated errors from \regularsys\, or cosmic variance. However, it is necessary to estimate the error introduced by \gpcr\, and determine whether it should be incorporated into the PNG parameter fitting.


\hui{Finally, in Section \ref{sec:ssi}, we discuss a  procedure to model 
\gpcr\, using real data. The process begins with an input `truth catalog', to which redshift and galaxy bias is assigned for each galaxy. We then generate an emulator based on source injection, producing a large number of synthetic galaxies. A selection function is subsequently applied to the output catalog. This approach yields the spatially varying redshift and galaxy bias distributions, which can be used to estimate the impact of 
\gpcr\, on 2-point or 3-point correlation functions.}


\section{Author Contribution Statements}
\noindent HK: Contributed to the conceptualization, code development, data analysis, and writing of the draft. 

\noindent NEC: Contributed to initial project planning, significant guidance and mentorship, and revision of the text. 

\noindent BL: Contributed significant guidance and mentorship, and revision of the text. 

\noindent EG: Contributed significant guidance and mentorship.

\noindent MRM: Contributed to general discussions.

\noindent NW: Contributed to general discussions.

\begin{acknowledgments}
This paper has undergone internal review by
the LSST Dark Energy Science Collaboration. The internal reviewers were Jack Elvin-Poole and Carlos Garcia-Garcia.
The authors thank the internal reviewers for their valuable comments.

We thank Alex Alarcon, Peter Ferguson, Ramon Miquel, Ashley Ross, Carles Sánchez, and Nikolina Šarčević for discussions and coding advice. 

We thank the anonymous reviewer for insightful suggestions that improved the manuscript.

This work makes use of the following software packages: \textsc{healpy} \footnote{\href{Healpy}{https://github.com/healpy/healpy}}, \textsc{ccl} \footnote{\href{CCL}{https://github.com/LSSTDESC/CCL/}}, \textsc{astropy} \footnote{\href{astropy}{https://www.astropy.org/}}, \textsc{jupyter} \footnote{\href{jupyter}{https://jupyter.org/}}, \textsc{namaster} \footnote{\href{NaMaster}{https://github.com/LSSTDESC/NaMaster}}, \textsc{camb} \footnote{\href{CAMB}{https://github.com/cmbant/camb}}, \textsc{matplotlib} \footnote{\href{matplotlib}{https://matplotlib.org}}, \textsc{scipy} \footnote{\href{scipy}{https://scipy.org/}}, \textsc{numpy} \footnote{\href{numpy}{https://matplotlib.org}}. 

The results lead in this paper have received funds from MCIN/AEI/10.13039/501100011033 and UE NextGenerationEU/PRTR (JDC2022-049551-I). 

IFAE is partially funded by the CERCA program of the Generalitat de Catalunya.

The DESC acknowledges ongoing support from the Institut National de 
Physique Nucl\'eaire et de Physique des Particules in France; the 
Science \& Technology Facilities Council in the United Kingdom; and the
Department of Energy and the LSST Discovery Alliance
in the United States.  DESC uses resources of the IN2P3 
Computing Center (CC-IN2P3--Lyon/Villeurbanne - France) funded by the 
Centre National de la Recherche Scientifique; the National Energy 
Research Scientific Computing Center, a DOE Office of Science User 
Facility supported by the Office of Science of the U.S.\ Department of
Energy under Contract No.\ DE-AC02-05CH11231; STFC DiRAC HPC Facilities, 
funded by UK BEIS National E-infrastructure capital grants; and the UK 
particle physics grid, supported by the GridPP Collaboration.  This 
work was performed in part under DOE Contract DE-AC02-76SF00515.
\end{acknowledgments}
\section{Data Availability}
The data that support the findings of this article are openly available \cite{SubSampleSystematics}, embargo periods may apply.
\appendix
\section{\texorpdfstring{$\langle f\delta,\delta \rangle$}{<f delta, delta>}, 
        \texorpdfstring{$\langle f\delta, f \rangle$}{<f delta, f>} \& 
        \texorpdfstring{$\langle f\delta, f\delta \rangle$}{<f delta, f delta>}}

\label{apdx:A}

For $\langle f\delta,\delta \rangle$ \& $\langle f\delta, f\rangle$ , the terms are in the format of 
\begin{equation}
     \mathrm{Corr}(AB,B) = \frac{E(ABB)-E(AB)E(B)}{std(AB)std(B)} 
\end{equation}
Here A and B are $f$ and $\delta$ or vice versa, As $f$ and $\delta$ are uncorrelated, the above equation can be written in the format 
\begin{equation}
    \mathrm{Corr}(AB,B) = \frac{E(A)E(BB)-E(AB)E(B)}{std(AB)std(B)} 
\end{equation}
Under our assumption, E(A) = E(B) = 0
\begin{equation}
    \mathrm{Corr}(AB,B) = 0
\end{equation}

These terms are always 0. 

For $\langle f\delta, f\delta\rangle$, it is in the format:
\begin{multline}
    \mathrm{Corr}(AB,AB) = \frac{E(AABB)-E(AB)E(AB)}{std(AB)std(AB)} \\
               = \frac{E(AA)^2E(BB)^2-E(A)^2E(B)^2}{std(A)^2std(B)^2}\\
               = \frac{E(AA)^2}{std(A)^2} \cdot \frac{E(BB)^2}{stdB)^2} \\
               = \mathrm{Corr}(A,A) \cdot \mathrm{Corr}(B,B)
\end{multline}
Thus, we have
\begin{equation}
    \langle f\delta, \delta\rangle = \langle f\delta, f\rangle = 0
\end{equation}

\begin{equation}
    \langle f\delta, f\delta\rangle = \langle f,f\rangle\langle \delta,\delta\rangle
    \label{equ:A6}
\end{equation}

\section{Weight-based systematics correction}
\label{apdx:B}
Weight-based systematics correction is the most widely used method for imaging systematics correction. It derives a sky-position dependent imaging systematics weight $w_{\mathrm{sys}}$, and apply this weight to galaxies or their associated randoms to correct the density fluctuation caused by imaging systematics. 

\begin{equation}
    \hat{\rho}_{\mathrm{corrected}} = w_{sys}\rho_{\mathrm{obs}}
\end{equation}

$\hat{\rho}_{\mathrm{corrected}}$ is the normalized corrected density field:
\begin{equation}
    \langle \hat{\rho}_{\mathrm{corrected}}\rangle=1
\end{equation}

If the density fluctuation for all sub-samples are the same and $w_{\mathrm{sys}}$
 is accurate, $\hat{\rho}_{\mathrm{corrected}}$ recovers the true galaxy density field. However, if different sub-samples have different systematics, the corrected galaxies density field is:

\begin{equation}
    \hat{\rho}_\mathrm{corrected} = w_{sys}\sum \rho_{\mathrm{truth},k}\prod_i(1+f_k^i)
\end{equation}

$f_k^i$ is defined in Equation \ref{equ:f_k-i}. $h_k$ is the fraction of galaxy sub-sample k. 

\begin{equation}
    \rho_{\mathrm{truth},k} = h_k \hat{\rho}_{\mathrm{truth},k}
\end{equation}

\begin{equation}
    \sum h_k = 1 
\end{equation}

\begin{equation}
    \hat{\rho}_{\mathrm{corrected}} = w_{sys} \sum_k \prod_i(1+f_k^i)h_k \hat{\rho}_{\mathrm{truth},k}
\end{equation}

Suppose that we have many realizations of true galaxies distributions (mocks),  and these mocks have the same imaging systematics. For any given sky position, we take the average of all mocks:

\begin{equation}
    \langle \hat{\rho}_{\mathrm{corrected}}\rangle = w_{sys} \sum_k \prod_i(1+f_k^i)h_k \langle \hat{\rho}_{\mathrm{truth},k}\rangle
\end{equation}

\begin{equation}
    \langle \hat{\rho}_{\mathrm{corrected}}\rangle = \langle \hat{\rho}_{\mathrm{truth},k}\rangle = 1
\end{equation}

Thus, the systematics weight has a relationship with the systematics trends with sub-samples as:

\begin{equation}
     w_{sys}  = \frac{1}{\sum_k \prod_i(1+f_k^i)h_k}
     \label{equ:B9}
\end{equation}

The density field for each corrected sub-sample is:

\begin{equation}
    \rho_{\mathrm{corrected},k} = \frac{\prod_i(1+f_k^i)}{\sum_k \prod_i(1+f_k^i)h_k} \rho_{\mathrm{truth},k}
\end{equation}

The systematics still exist for the sub-samples even after an ideal systematics correction. 

Next, we compute the correlation function for this corrected galaxy density field: 

\begin{multline}
    \langle \delta_{\mathrm{corrected}}, \delta_{\mathrm{corrected}} \rangle = \\
    \langle w_{\mathrm{sys}}\sum_k h_k(\delta_{\mathrm{truth},k}+1)\prod_i(1+f_k^i) - 1, \\
    w_{\mathrm{sys}}\sum_{k'} h_{k'}(\delta_{\mathrm{truth},k'}+1)\prod_i(1+f_k^i) - 1 \rangle 
\end{multline}

This equation can be decomposed into several terms:

\begin{multline}
    \langle \delta_{\mathrm{corrected}}, \delta_{\mathrm{corrected}} \rangle = \\
    \langle w_{\mathrm{sys}} \sum_k h_k \prod_i(1+f_k^i) \delta_{\mathrm{truth}, k}, w_{\mathrm{sys}}\sum_{k} h_{k} \prod_i(1+f_k^i) \delta_{\mathrm{truth}, k} \rangle \\
    + \langle w_{\mathrm{sys}} \sum_k h_k \prod_i(f_k^i + 1) -1, w_{\mathrm{sys}} \sum_k h_k \prod_i(f_k^i + 1) -1 \rangle
    \label{equ:B12}
\end{multline}

According to equation \ref{equ:B9}, the second term is 0. This is the 1st order systematics term. It vanishes when $w_{\mathrm{sys}}$ is accurate. We define a new term:

\begin{equation}
    \mu_k = 
    \frac{\prod_i(1+f_k^i)}{\sum_k h_k\prod_i(1+f_k^i)} = 
    \frac{w_\mathrm{sys}}{w_\mathrm{sys, k}}
\end{equation}

$\mu_k$ can also be seen as the global systematics weight $w_\mathrm{sys}$ divided by the accurate systematics weight applied to sub-sample-$k$ $w_\mathrm{sys,k}$. $\mu_k$ is internally related to each other:

\begin{equation}
    \sum_k h_k(\mu_k-1) = 0
    \label{equ:B14}
\end{equation}

According to equation \ref{equ:A6}, un-correlated terms can be separated. Equation \ref{equ:B12} is:

\begin{multline}
    \langle \delta_{\mathrm{corrected}}, \delta_{\mathrm{corrected}} \rangle = \\
    \sum_{k,k'} h_k h_{k'} \langle \mu_k, \mu_{k'} \rangle \langle \delta_{\mathrm{truth},k}, \delta_{\mathrm{truth}, k'} \rangle
\end{multline}

We further isolate the true clustering signal: 

\begin{multline}
    \langle \delta_{\mathrm{corrected}}, \delta_{\mathrm{corrected}} \rangle = 
    \langle \delta_\mathrm{truth}, \delta_\mathrm{truth} \rangle +
    \\
    \sum_{k,k'} h_k h_{k'} \langle \mu_k-1, \mu_{k'}-1 \rangle \langle \delta_{\mathrm{truth},k}, \delta_{\mathrm{truth}, k'} \rangle
\end{multline}

We define:
\begin{equation}
    f_k^\mathrm{corrected} = \mu_k - 1
    \label{equ:fcorrected}
\end{equation}

$f_k^\mathrm{corrected}$ is the systematics trend of galaxy sub-sample $k$ after an optimal imaging systematics weight correction. In conclusion, the presence of sub-sample systematics trend modifies the corrected galaxy correlation function, and it can be parametrized with $f_k^\mathrm{corrected}$.

\section{Correlation-function-based systematics correction}
\label{apdx:C}

Correlation-function-based systematics correction applies imaging systematics correction on correlation-function level. \hui{Standard methods\cite{ho2012clustering} in this category correct the correlation functions to 1st order. Extensions of such methods\cite{shafer2015multiplicative} further corrects multiplicative errors in 2nd order. }

\hui{In this section, we show that if we care about systematics to 1st order, the correction methods are still valid even if we assume that sub-sample systematics exists. However, if we want to correct for 2nd order terms, the currently available methods are not sufficient enough to correct sub-sample systematics. }

The observed correlation function can be expressed as: 

\begin{multline}
    \langle \delta_{\mathrm{obs}}, \delta_{\mathrm{obs}} \rangle = \\
    \langle \delta_{\mathrm{truth}}, \delta_{\mathrm{truth}} \rangle +  \sum_{i,j} \langle f^{i}, f^{j} \rangle + \sum_{i,j} \langle f^{i} \delta_{\mathrm{truth}}, f^{j} \delta_{\mathrm{truth}} \rangle + ...
    \label{equ:C1}
\end{multline}

We use superscript for $f^{i}$, $f^{j}$ to reflect different types of systematics, e.g. depth, stellar density, etc. These methods relies on the fact that cross-correlating $f^{i}$ with the observed galaxy density field gives the second term:

\begin{equation}
    \langle f^{i}, \delta_{\mathrm{obs}} \rangle = \sum_j \langle f^i, f^j \rangle 
\end{equation}

The basic correction model ignores the 3rd terms and beyond, and assumes linear systematics\cite{elvin2018dark, ross2011ameliorating, kalus2019map, alonso2019unified, nicola2020tomographic}: 

\begin{equation}
    f^i(\textbf{sys}^i) = a \cdot \textbf{sys}^i + b
    \label{equ:linear-sys}
\end{equation}

Here $\mathrm{sys}_i$ are the systematics values in survey property maps. Under these assumptions, the true galaxy density field $\delta_{\mathrm{truth}}$ can be strictly solved. Several methods have been developed for such problem, and \cite{weaverdyck2021mitigating} proved that these methods are mathematically equivalent. For convenience, we only choose expression from Template Subtraction (TS,\cite{ho2012clustering}) for further discussion. Within the TS context, the observed galaxy density field fluctuation is modified by imaging systematics map $i$ with: 
\begin{equation}
    \delta_{\mathrm{obs}} = \delta_{\mathrm{TS}} + \sum_i \epsilon^i \cdot \textbf{sys}^i
\end{equation}

$\textbf{sys}^i$ are the map values for a given systematics map, and its mean is shifted to 0. $\delta_\mathrm{TS}$ is the corrected galaxy density field which is very similar to the truth. $\epsilon^i$ is a constant value. Measuring the correlation function of $\delta_{\mathrm{obs}}$ gives us: 

\begin{equation}
\label{equ:TS-Cl}
C_{\ell}^{\mathrm{obs}} = C_{\ell}^{\mathrm{TS}} + \sum_{i,j}^{n} \epsilon^i \epsilon^j C_{\ell}^{ij}
\end{equation}

Here $C_{\ell}^{\mathrm{obs}}$ is the observed power spectrum, $C_{\ell}^{\mathrm{TS}}$ is the TS-corrected power spectrum, and $C_{\ell}^{ij}$ is the cross power spectrum between two survey property maps, $n$ is the total number of used maps. $\epsilon^i$ can be decomposed into different $\ell$ modes, but it is typically considered as an $\ell$-independent parameter. The coefficients $\epsilon^i$ are solved with a set of equations

\begin{equation}
C_{\ell}^{\mathrm{obs},i} = \sum_{j}^{n} \epsilon^j C_{\ell}^{ij}
\end{equation}

\noindent $C_{\ell}^{\mathrm{obs},i}$ is the cross-power spectrum between observed galaxy density and survey property maps. $C_{\ell}^{TS}$ need to be further de-biased\cite{elsner2016unbiased} to get the true power spectrum $C_{\ell}^{\mathrm{truth}}$:  

\begin{equation}
C_{\ell}^{\mathrm{truth}} \sim C_{\ell}^{TS} \left (1 - f^2_{\mathrm{sky}}\frac{n}{2\ell+1}\right)
\end{equation}
Here $f_{\mathrm{sky}}$ is the fractional area for this survey. $n$ is still the total number of used maps. The solution with $\epsilon^i$ depends on $\ell$. For linear imaging systematics mitigation, the mean of $\epsilon^i$ across all $\ell$ is chosen. Such a procedure avoids over-fitting the true cosmological fluctuations.

\subsection{Template Subtraction with multiple sub-samples}
\label{sec:linear-response}
If all galaxies respond to the survey property maps linearly, 

\begin{equation}
\delta_{\mathrm{obs}, k} = \delta_{\mathrm{TS}, k} + \sum_i \epsilon_k^i \cdot \textbf{sys}^i
\end{equation}

If we sum over all galaxy sub-samples $k$, and compute the power spectrum, we have

\begin{multline}
\langle \sum_k \delta_{\mathrm{obs}, k}, \sum_k \delta_{\mathrm{obs}, k} \rangle = \\
\langle \sum_k h_k\delta_{\mathrm{TS}, k}+\sum_{ki} h_k\epsilon_k^i \cdot \textbf{sys}^i, \sum_i h_k\delta_{\mathrm{TS}, k} + \sum_{ki} h_k\epsilon_k^i \cdot \textbf{sys}^i  \rangle
\end{multline}

$h_k$ is defined in Equation \ref{equ:fraction_hk}. $\sum_k \delta_{\mathrm{obs}, k}$ is the observed galaxy density of the whole sample $\delta_\mathrm{obs}$. The above equation can be simplified as 

\begin{multline}
\label{equ:multi-TS}
\left\langle \delta_\mathrm{obs}  \delta_{\mathrm{obs}} \right\rangle =\\
\left\langle \delta_\mathrm{TS}  \delta_\mathrm{TS} \right\rangle + \sum_{ij} \left[ \left\langle \textbf{sys}^i, \textbf{sys}^{j} \right\rangle \sum_{k} h_k\epsilon_k^i \cdot \sum_{k'}h_{k'}\epsilon_{k'}^{j} \right]
\end{multline}

If we define a new coefficient 
\begin{equation}
\epsilon^i = \sum_{k} h_k\epsilon_k^i 
\label{equ:C11}
\end{equation}

Then equation \ref{equ:multi-TS} can be transformed into the format for deriving the TS coefficient in equation \ref{equ:TS-Cl}. Therefore, if all types of galaxies in a sample scale linearly with survey properly maps, then the validity of TS still holds, even if their response coefficient $\epsilon_i^k$ is different. After applying linear imaging systematics weight to the galaxy sample as a whole, the slope of each sub-type galaxy and their densities have a certain relationship:

\begin{equation}
\sum_{k} h_{k} \epsilon_{k, \mathrm{corrected}}^i= 0
\label{equ:linear-internal}
\end{equation}

In fact, this is a variant of equation \ref{equ:B14}. In conclusion, if we ignore higher order terms in equation \ref{equ:C1}, \gpcr\, is also ignored as it belongs to higher order terms. 

\subsection{Multiplicative Error} 

Some methods also consider higher order terms in equation \ref{equ:C1}. They are defined as multiplicative error. When considering one term higher, equation \ref{equ:C1} becomes:

\begin{multline}
    \langle \delta^\mathrm{obs}, \delta^\mathrm{obs} \rangle = \\ 
    \langle \delta_\ell^\mathrm{truth},  \delta_\ell^\mathrm{truth} \rangle + \sum_{i,j}\epsilon^i \epsilon^j \langle \textbf{sys}^i, \textbf{sys}^j \rangle + \\ 
    \langle \delta^\mathrm{truth}, \delta^\mathrm{truth}\rangle \sum_{ij} \epsilon^i\epsilon^j \langle \textbf{sys}^i, \textbf{sys}^j \rangle 
    \label{equ:C13}
\end{multline}

We cannot decouple this equation into distinct $\ell$ modes due to the presence of the last term. We can replace the $C_\ell$s in the standard TS method with the angular correlation function $w(\theta)$ \cite{ross2011ameliorating}.  Solving this equation is non-trivial, but still possible. \cite{shafer2015multiplicative} and \cite{weaverdyck2021mitigating} present methods to correct for the multiplicative error, and yield accurate estimates of $\epsilon^i$.

If we consider galaxy sub-samples, the last term in equation \ref{equ:C13} becomes: 

\begin{equation}
    \sum_{ij} \left ( \langle \textbf{sys}^i, \textbf{sys}^j \rangle \sum_{kk'} \epsilon^i_k \epsilon^j_{k'}  \langle \delta^{\mathrm{truth}}_{\mathcal{M}_k}, \delta^{\mathrm{truth}}_{\mathcal{M}_{k'}} \rangle \right )
\end{equation}

Unlike equation \ref{equ:multi-TS},
the subscript $k$, $k'$ here also exists inside the correlation function for sub-sample $\mathcal{M}_k$ and $\mathcal{M}_{k'}$. Thus, when considering higher order terms, \gpcr\, can not be transformed into equation \ref{equ:C13}. The correction methods that works for multiplicative error can not be applied to correct for \gpcr. 

In conclusion, the basic correction-function-based systematics correction methods correct the 1st order terms, leaving un-corrected 2nd order terms and beyond. By defining a new coefficient in Equation \ref{equ:C11}, \gpcr\, can be transformed into the original mathematical model. Thus, if 1st order accuracy is sufficient, it is unnecessary to consider \gpcr. If we further consider 2nd-order terms, \gpcr\, can not be mathematically transformed into the multiplicative error model.

\section{Impact of \gpcr\, with different systematics map patterns}
\label{apdx:D}

\gpcr\, is a non-linear effect. Even if two different systematics maps cause the same level of galaxy density fluctuation, their impact on the observed correlation function can be every different. In this section, we use a toy model to visualize the impact of \gpcr\, on a corrected sample with internal linear variation described by Equation \ref{equ:linear-internal}.

Suppose we have two galaxy samples produced by an underlying power spectrum $C_\ell$, and they are uncorrelated. They both have linear systematics trends against a systematics map. We perform a linear transformation to the systematics map to give a mean of 0 and a minimum of -1. We define this transformed map as $f$. The two galaxy samples are modified by $f$ in an inverse direction:

\begin{equation}
    \delta_{\mathrm{obs},\mathcal{M}_1} = \delta_{\mathrm{truth},\mathcal{M}_1} + f+ f \cdot \delta_{\mathrm{truth},\mathcal{M}_1}
\end{equation}

\begin{equation}
    \delta_{\mathrm{obs},\mathcal{M}_2} = \delta_{\mathrm{truth},\mathcal{M}_2} - f - f \cdot \delta_{\mathrm{truth},\mathcal{M}_2}
\end{equation}

The sum of the two fields is

\begin{equation}
    \delta_{\mathrm{obs},\mathcal{M}_{12}} = \delta_{\mathrm{truth},\mathcal{M}_{12}}+f\cdot (\delta_{\mathrm{truth},\mathcal{M}_1}-\delta_{\mathrm{truth},\mathcal{M}_2})
\end{equation}

We suppose that the observer can only see  $\delta_{\mathrm{obs},\mathcal{M}_{12}}$ and do not know that there are two galaxy samples internally varying. The observer can only check the \ngal\, trend against $\delta_{\mathrm{obs},\mathcal{M}_{12}}$ and conclude that there is no systematics in this sample. 

The correlation function of $\delta_{\mathrm{obs},\mathcal{M}_{12}}$ is

\begin{multline}
    \langle\delta_{\mathrm{obs},\mathcal{M}_{12}},\delta_{\mathrm{obs},\mathcal{M}_{12}}\rangle  = \langle\delta_{\mathrm{truth}, \mathcal{M}_{12}},\delta_{\mathrm{truth}, \mathcal{M}_{12}}\rangle  + \\
    \langle f(\delta_{\mathrm{truth}, \mathcal{M}_{1}} -\delta_{\mathrm{truth},\mathcal{M}_{2}} ), f(\delta_{\mathrm{truth}, \mathcal{M}_{1}} -\delta_{\mathrm{truth}, \mathcal{M}_{2}} )\rangle 
\end{multline}

The second term is non-linear and can not be trivially decomposed. We used a toy model to demonstrate its impact on the power spectrum.

\begin{figure}
    \centering
    \includegraphics[width=\linewidth]{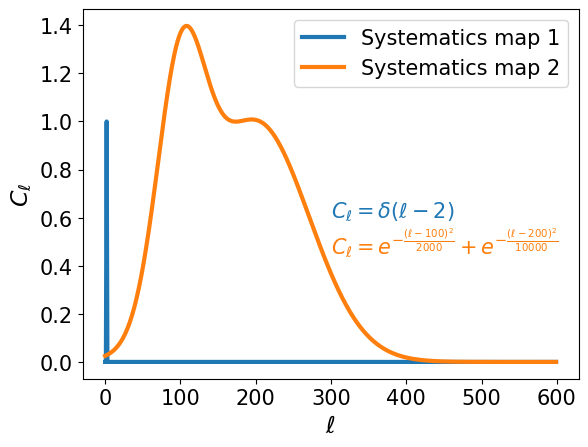}
    \caption{The power spectrum of systematics map. The observed galaxy densities for each sub-sample is linearly modified by the maps while the combined whole sample is uncorrelated with the maps.}
    \label{fig:ApdxA1}
\end{figure}

\begin{figure}
    \centering
    \includegraphics[width=\linewidth]{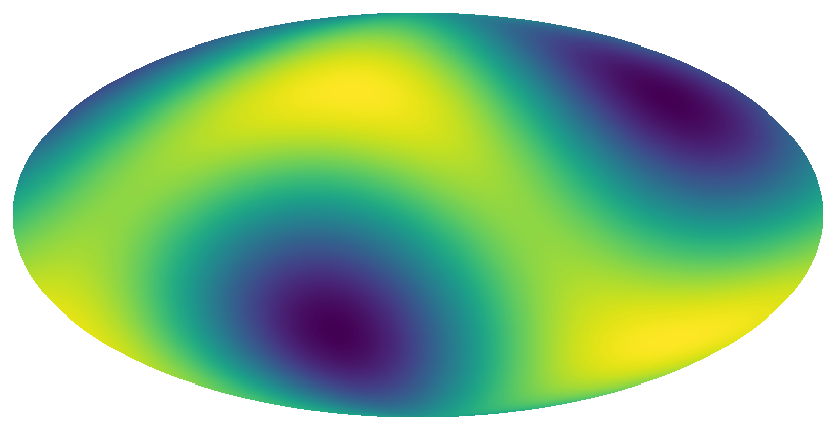}
    \caption{A fiducial imaging systematics map with large-scale patterns only. }
    \label{fig:ApdxA1a}
\end{figure}

\begin{figure}
    \centering
    \includegraphics[width=\linewidth]{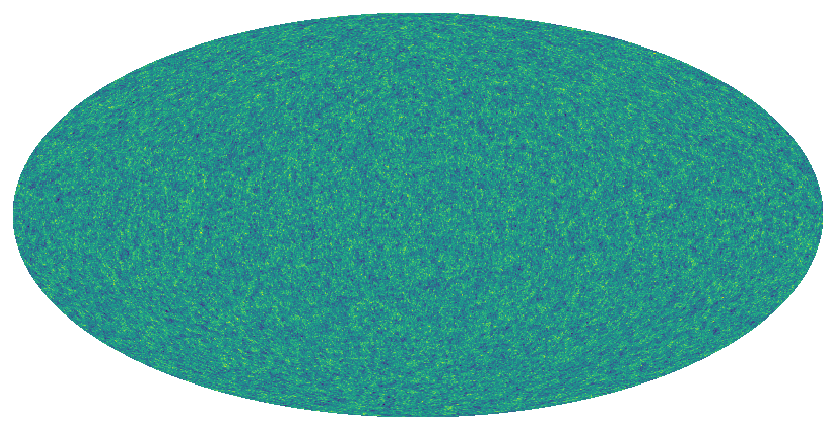}
    \caption{A fiducial imaging systematics map with small-scale patterns only.}
    \label{fig:ApdxA1b}
\end{figure}

We use a fiducial $C_\ell$ to produce two uncorrelated galaxy fields. We test on two systematics maps. Systematics map 1 only has a large-scale mode at $\ell=2$, while systematics map 2 has small-scale modes. The power spectrum of the two systematics maps are in Figure \ref{fig:ApdxA1}. Their patterns are shown in Figure \ref{fig:ApdxA1a} and \ref{fig:ApdxA1b}. The numbers in the maps directly determine how $\delta_{\mathrm{truth}}$ is modified in each given pixel. We assume large systematics variations with amplitude $\approx 100\%$. This is not realistic for an actual galaxy sample. Thus, the absolute number derived in this section is not important. We care more about how different patterns can produce different amplitude shifts in the observed $C_\ell$. 

\begin{figure}
    \centering
    \includegraphics[width = \linewidth]{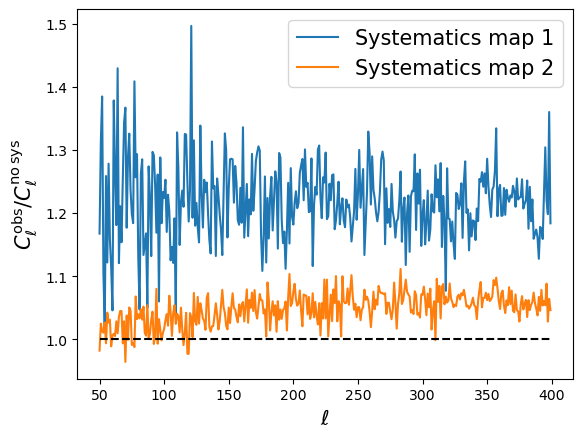}
    \caption{A combined sample of tracers with galaxy bias $b_1$, $b_2$. In the blue curve, the two tracers are uncorrelated with the contaminated map \textbf{X}. In the orange curve, the galaxy density of the two tracers are linearly modified by map \textbf{X}. However, the combined sample is uncorrelated with \textbf{X}. }
    \label{fig:ApdxA2}
\end{figure}

Figure \ref{fig:ApdxA2} shows the ratio of the observed $C_\ell$ and the $C_\ell$ without imaging systematics (sample $\mathcal{M}_1$, $\mathcal{M}_2$ do not have internal variation). We see that systematics map 1 produces a larger change in amplitude than systematics map 2. At an angular separation much smaller than the systematics variation scale, the central galaxy in map 1 is more likely to find galaxy pairs similar to itself.

\section{Mock Validation}
\label{apdx:E}
\subsection{shift with galaxy bias}

\label{sec:bshift-mock}

We test the equation derived in Section \ref{sec:shift_bias} by producing a mock field according to Equation \ref{equ:b-shift-field}. We generate a contaminated field by a custom-defined equation:

\begin{multline}
    C_{\ell,\mathrm{sys}} = e^{-\frac{(l-15)^2}{200}} + e^{-\frac{(l-50)^2}{2000}} + \\
    0.5e^{-\frac{(l-410)^2}{3000}} + 0.5e^{-\frac{(l-200)^2}{10000}}
\end{multline}

We generate a \textsc{healpix} density field with $C_{\ell, \mathrm{sys}}$, and assign tracer 1 to be in the region where the pixel value is greater than 50, comprising 27\% of the total area, and tracer 2 is in the rest (73\%) of the region. 

We populate a matter density field with a power spectrum generated with 
\textsc{pyccl} \cite{chisari2019core}, and mutiply this field by 1, 2, and 10 to produce a biased galaxy density field with bias = 1,2 and 10. We consider two senarios. In the first senario, tracer 1 has galaxy bias $b_1$= 1, tracer 2 has galaxy bias $b_2$ = 2. In the second senairio, $b_1$ = 1, $b_2$ = 10. The two tracers are non-overlapping and such model defines a condition of `extreme galaxy clumping': $\Delta w$ defined in equation \ref{equ:shift-bias-final} is small. The amplitude of the observed correlation function is close to $b_{eff,v2}$ in Equation  \ref{equ:b-shift-bias-v2} rather than $b_{eff,v1}$ in Equation \ref{equ:b-shift-bias-v1}.

\begin{figure}
    \centering
    \includegraphics[width=\linewidth]{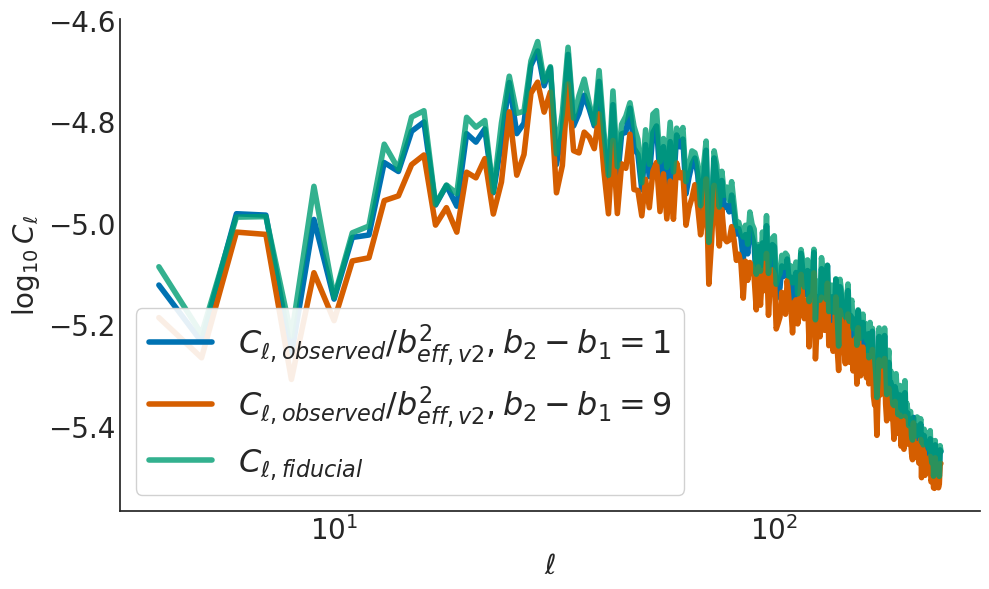}
    \caption{Power spectrum $C_\ell$ of two correlated galaxy tracers in the same redshift range, but with different galaxy bias. Their distribition is modified by Equation \ref{equ:b-shift-field}. The observed power spectrum is divided by the effective galaxy bias $b_{eff,v2}$ in Equation \ref{equ:b-shift-bias-v2}. The green curve is the fiducial $C_\ell$ which corresponds to $w_m$ in Equation \ref{equ:wm}. The blue curve is the observed $C_\ell$ for two tracers with galaxy bias $b_1$ = 1, $b_2$ =2. Similarly, the orange curve is for two tracers with galaxy bias $b_1$=1. $b_2$=10.}
    \label{fig:bshift-pk}
\end{figure}

\begin{figure}[htbp]
    \centering
    \includegraphics[width=\linewidth]{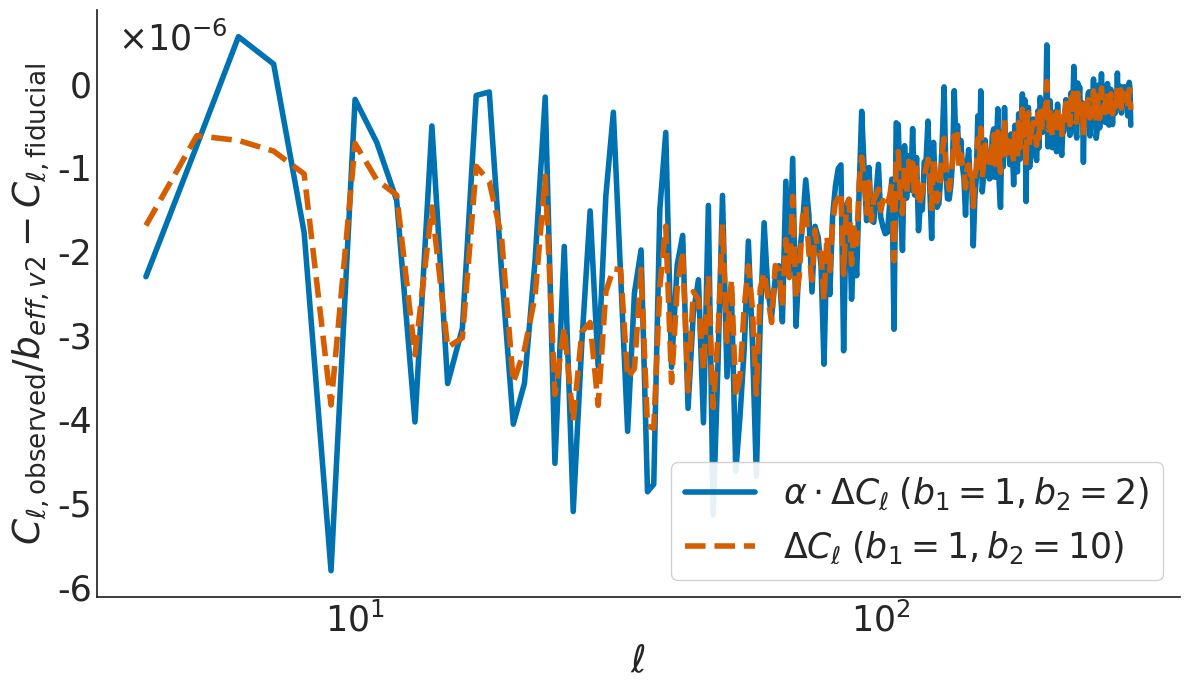}
    \caption{Difference between the observed and the fiducial $C_\ell$. The observed $C_\ell$ is divided by the effective galaxy bias in Equation \ref{equ:b-shift-bias-v2}. This property is closely related to the `artificial window function' in Equation \ref{equ:shift-bias-final}. The amplitude of $\Delta C_\ell$ is controlled by the galaxy bias of the two samples. For galaxy pairs with difference galaxy bias, their difference can be analytically computed with Equation \ref{equ:alpha}. In this toy model, the difference between the two galaxy bias is $\alpha$ $\sim$ 3.5. }
    \label{fig:bshift-cl-diff}
\end{figure}

Figure \ref{fig:bshift-pk} shows the power spectrum of these two senarios divided by $b_{eff,v2}$, compared with the power spectrum from the fiducial power spectrum $C_{\ell, fiducial}$, which is the power spectrum of $\delta_m$. Figure \ref{fig:bshift-cl-diff} shows the difference of the observed power spectrum divided by $b_{eff,v2}$, and the fiducial power spectrum. This difference is determined by the `artificial window function'. The amplitude is controlled by the galaxy bias. 
\begin{equation}
    \alpha_i = \frac{(b_1-b_2)^2}{b_{eff,v2}^2}
    \label{equ:alpha_i}
\end{equation}
We get different $\alpha_i$ for different $b_1$, $b_2$ pairs. When comparing the amplitude between the $b_1$=1, $b_2$=2 and the $b_1$=1, $b_2$=10 pair, we can measure $\alpha_1$, $\alpha_2$ with equation \ref{equ:alpha_i}. And their amplitude has a ratio of 
\begin{equation}
    \alpha = \alpha_1/\alpha_2
    \label{equ:alpha}
\end{equation}

\subsection{Shift in redshift}
\label{sec:zshift-mock}
\begin{figure}[htbp]
    \centering
    \includegraphics[width=\linewidth]{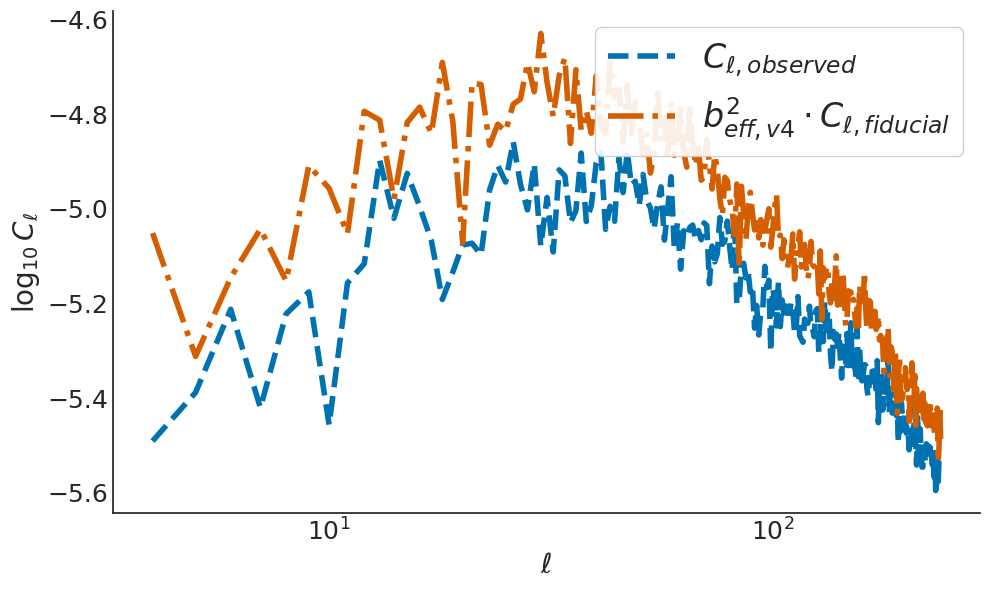}
    \caption{Observed power spectrum of two uncorrelated galaxy tracers with a spatial distribution defined in Equation~\ref{equ:delta_obs_z_shift}. The orange curve shows a fiducial power spectrum with amplitude $b_{\mathrm{eff},v4}$ in Equation~\ref{equ:b_eff_v4}. The blue curve shows the observed power spectrum defined in Equation~\ref{equ:w_obs_z_shift_clumping}.}
    \label{fig:zshift-pk}
\end{figure}

We test the equations derived in Section \ref{sec:shift_in_z} with a toy mock. We adopt the same contamination model and galaxy power spectrum as in Section \ref{sec:bshift-mock}. We produce two uncorrelated density field with the same input power spectrum. Figure \ref{fig:zshift-pk} shows fiducial and the observed power spectrum. Indeed, as we discussed in Section \ref{sec:shift_in_z}, the observed power and the fiducial power spectrum has a larger difference compared with the correlated tracers discussed in \ref{sec:bshift-mock}: A result of $|r_z| > |r_b|$ in Equation \ref{equ:rb} and \ref{equ:rz}. 

\subsection{Discussion}
That the window function for the `shift in redshift' case has a larger impact than the `shift in galaxy bias' case discussed in section \ref{sec:shift_bias}. The first term divided by the second term here is  
\begin{equation}
    r_{z} = -h_{1}\Delta Win_1 \cdot \frac{b_1^2+b_2^2}{f_1 b_1^2 + f_2 b_2^2}
    \label{equ:rz}
\end{equation}

while for equation \ref{equ:shift-bias-final}, it is 
\begin{equation}
r_{b} = -h_{1}\Delta Win_1 \cdot \frac{(b_{1} - b_{2})^2}{f_{1}b_{1}^2+f_{2}b_{2}^2}
\label{equ:rb}
\end{equation}

$|r_z|>|r_b|$, meaning that this impact is always larger when the galaxy sample has a shift in redshift, rather than a shift in galaxy bias. This is also conceptually true. When there is a shift in redshift, the cross-correlation signals are all lost for the two tracers. When there is a shift in galaxy bias, the cross-correlation signal still partially remains. Since $r_b$ and $r_z$ scales with the `hypothetical window function' $h_1\Delta Win_1$, the model of two uncorrelated tracers is more sensitive to this window function. 

\clearpage

\bibliography{apssamp}
\end{document}